\documentclass[prd,aps,showpacs,floats,letterpaper,floatfix,
groupedaddress,superscriptaddress
,nofootinbib
,eqsecnum,
twocolumn
]{revtex4}
\usepackage{amsmath}
\usepackage{amsfonts}
\usepackage{bm}
\usepackage{graphicx}

\def\v1v2{{\bf v}_1 \cdot {\bf v}_2}
\def\tp{\tilde{p}}
\def\te{\tilde{e}}
\def\tal{\tilde{\alpha}}
\def\tbe{\tilde{\beta}}
\def\tom{\tilde{\omega}}
\def\tOm{\tilde{\Omega}}
\def\ti{\tilde{\iota}}

\allowdisplaybreaks

\begin{document}

\title{Relativistic orbits around spinning supermassive black holes.  Secular evolution to 4.5 post-Newtonian order}

\author{
Clifford M.~Will} \email{cmw@phys.ufl.edu}
\affiliation{Department of Physics, University of Florida, Gainesville, Florida 32611, USA}
\affiliation{GReCO, Institut d'Astrophysique de Paris, CNRS,\\ 
Universit\'e Pierre et Marie Curie, 98 bis Bd. Arago, 75014 Paris, France}
\author{
Matthew Maitra} \email{mam221@cam.ac.uk}
\affiliation{Trinity College, University of Cambridge, Cambridge CB2 1TQ UK }

\date{\today}

\begin{abstract}
We derive the secular evolution of the orbital elements of a stellar-mass object orbiting a spinning massive black hole.   We use the post-Newtonian approximation in harmonic coordinates, with test-body equations of motion for the conservative dynamics that are valid through 3PN order, including spin-orbit, quadrupole and (spin)$^2$ effects, and with radiation-reaction contributions linear in the mass of the body that are valid through 4.5PN order, including the 4PN damping effects of spin-orbit coupling.  The evolution equations for the osculating orbit elements are iterated to high PN orders using a two-timescale approach and averaging over orbital timescales.   We derive a criterion for terminating the orbit when its Carter constant drops below a critical value, whereupon the body plunges across the event horizon at the next closest approach.  The results are valid for arbitrary eccentricities and arbitrary inclinations.  We then analyze numerically the orbits of objects injected into high-eccentricity orbits via interactions within a surrounding star cluster, obtaining the number of orbits and the elapsed time between injection and plunge, and the residual orbital eccentricity at plunge as a function of inclination.  We derive an analytic approximation for the time to plunge in terms of initial orbital variables.    We show that, if the black hole is spinning rapidly, the flux of gravitational radiation during the final orbit before plunge may be suppressed by as much as three orders of magnitude if the orbit is retrograde on the equatorial plane compared to its prograde counterpart.  
\end{abstract}

%\pacs{}
\maketitle

\section{Introduction and summary}
\label{sec:intro}

The relativistic motion of a star or small black hole in the field of a spinning massive black hole is a subject of much current interest, on two fronts.  Such orbits may lead to gravitational-wave (GW) emission at a level that is potentially detectable by ground-based laser-interferometric detectors, such as LIGO-VIRGO, or by a future space-based detector, such as LISA \cite{2014arXiv1411.5253B}.   In addition, relativistic effects on the orbits may play an important role in the evolution of dense nuclear clusters of stars and black holes orbiting a massive black hole at the center of a galaxy (for a recent review, see \cite{alexanderARAA}).  Indeed, these two problems are intertwined: interactions among the bodies in the cluster may inject bodies into orbits that pass very close to the black hole, leading to the emission of waves and a consequent inspiral into the hole.   

At the one extreme -- very close to the massive black hole -- the analysis of gravitational-wave emission and inspiral is a notoriously difficult problem, particularly if one seeks sufficiently accurate predictions to be used effectively in the analysis of data from gravitational-wave detectors.   A leading approach to this problem is the ``self-force'' program, which attempts to go beyond simple black-hole perturbation theory by incorporating the back-reaction effects on the geometry of the small mass of the orbiting particle, while keeping the background Kerr geometry of the black hole exact (see \cite{2011LRR....14....7P} for a review).  While great progress has been made, concrete results are currently limited to orbits around Schwarzschild black holes and equatorial orbits around Kerr black holes (see, {\em e.g.} \cite{2016PhRvD..94d4034V}).   Other approaches are hybrid in nature, combining information on the flux of radiation from black-hole perturbation theory with assumptions of energy and angular momentum balance, separations of timescales, and adiabatic evolution of orbital parameters \cite{2002PhRvD..66f4005G,2006PhRvD..73f4037G,2008PhRvD..78f4028H,2016arXiv160102042H}.   

At the other extreme -- very far from the black hole -- relativistic effects are small, but can have important long-term effects on the evolution of the cluster.  A leading example of this is the so-called ``Schwarzschild barrier'', in which the relativistic advance of the pericenter of a given orbit can, over long timescales, reduce the probability of the star being injected into an orbit that becomes a gravitational-wave driven, ``extreme mass-ratio'' inspiral (EMRI) \cite{2006ApJ...645.1152H,mamw2,2014MNRAS.443..355H,2014CQGra..31x4003B,2014MNRAS.437.1259B,2016ApJ...820..129B}.   In this realm, post-Newtonian (PN) theory, the weak-field, slow-motion approximation to general relativity, is adequate for including the dominant relativistic effects.  These can include the standard PN effects of the Schwarzschild geometry, notably the pericenter advance, as well as the effects of frame-dragging and the quadrupole moment of the Kerr geometry, and post-Newtonian ``cross-terms'' \cite{2014PhRvD..89d4043W} arising from the non-linear interactions between the field of the black hole and the fields of the orbiting bodies.  

Our interest lies between these extremes.  We wish to use post-Newtonian theory to study  a star or black hole that has just been injected into an orbit of such low angular momentum that it will pass close to the black hole, and thus will have its motion strongly influenced by gravitational radiation.  Ignoring further perturbations by other stars, we will then follow its motion until its orbital parameters require the body to cross the event horizon at the next closest approach.    To get a sense of what this means in terms of orbital parameters, consider a $10^6 M_\odot$ black hole, surrounded by a cluster of stellar-mass objects at a distance of a few milliparsecs.  The Schwarzschild radius of such a black hole is 50 nanoparsecs.   Thus, in order to get close to the black hole, the body must be injected into a very low angular momentum orbit, $L \sim [GMa(1-e^2)]^{1/2}  \sim bv$ where $a$ is the semi-major axis, of order a few milliparsecs, $e$ is the eccentricity, $b$ is the impact parameter, which has the scale of Schwarzschild radii, and $v \sim (GM/a)^{1/2}$ is the orbital velocity.   This gives $(1-e^2) \sim (b/a)^2$.  For $b \sim 100$ Schwarzschild radii, and $a \sim 1$ mpc,  this implies $1-e \sim 10^{-5}$.   Such extreme eccentricities are not the usual realm for applications of PN theory.   

Still, initially the star is not in the final, highly relativistic inspiral regime, and thus we might expect PN theory to be a reasonable approach to the study of such orbits, particularly if it is carried out to high orders in the PN expansion.   The PN approximation has proven to be ``unreasonably effective'' \cite{2011PNAS..108.5938W} in describing the quasi-circular inspiral of comparable mass-compact bodies, such as black holes or neutron stars.   In regimes where it might be expected to fail, it agrees surprisingly well with results from numerical relativity.    Whether this effectiveness will carry over to the highly eccentric, exreme mass-ratio inspiral problem will depend in part on what questions are being addressed.    If the question is ``what is the precise, late-time gravitational waveform including phase evolutions'', then there is reason to be skeptical of any answer based on PN theory, absent some corroboration from self-force theory, numerical relativity or some other technique.  However if the questions include
\begin{itemize}
\item
how many orbits or how much time does it take to go from  initial injection to plunge into the black hole?

\item
what is the distribution of late-time orbital eccentricity as a function of inclination of the orbit relative to the black hole's equatorial plane?

\item
how do the late-time energy flux and gravitational-wave frequency depend on inclination?

\end{itemize}
then PN theory might give quantitative estimates.   Time to plunge, for example, is useful input into schemes for evolving the cluster, including the effects of the loss of stars to the inspiral process.  Whether the late-time orbits are highly circularized or could have significant residual eccentricities, depending on inclination, could be important guides for research on such schemes as the self-force program.   And the dependence of the gravitational-wave signal on inclination could impact estimates of the detectability of EMRI radiation by space detectors such as LISA.
Questions such as these will be the main focus of this paper.  

We first develop the necessary analytic formulae to describe the evolution of very eccentric orbits from injection to plunge, for a stellar-mass object of mass $m$ in the gravitational field of a massive Kerr black hole of mass $M$ and dimensionless spin parameter $\chi$, derived to  high orders in the PN approximation. 
In Sec.\ \ref{sec:eom} we derive and display the basic two-body equations of motion to be used. They are valid through 3PN order for the conservative part of the orbital dynamics of a test particle, including frame-dragging, quadrupole-moment and (spin)$^2$ effects.  They are valid through 4.5PN order in the dissipative part caused by gravitational radiation reaction, including the 4PN-order effects of spin-orbit coupling.  The key results are Eqs.\ (\ref{eq:eom2}) and (\ref{eq:eomRR}) (together with detailed formulae displayed in Appendix \ref{app:Kerr}), expressed in harmonic coordinates.  The dissipative terms are given to first order in the reduced mass parameter $\eta = mM/(m+M)^2$, which is assumed to be very small.  They are presented in a ready-to-use form that permits them to be integrated into an $N$-body code for evolving the stellar cluster.

Section \ref{sec:lagrange} uses the formalism of osculating orbit elements and the ``Lagrange planetary equations'' for evolving the orbit elements in the presence of the relativistic, non-Keplerian perturbations (Sec.\ \ref{sec:basiceq}).   We use a two-scale analysis to separate each orbit element into an orbit-averaged piece and a piece that varies on an orbital timescale, and derive equations both  for the ``secular'' variations of the average elements on longer timescales, and for the periodic variations (Sec.\ \ref{sec:twoscale} and Appendix \ref{app:twoscale}).   Because we are working to high orders in a PN expansion, we must carefully iterate the planetary equations to high orders.   From the conservative sector of the equations of motion (Sec.\ \ref{sec:conservative}), we obtain the usual advance of the pericenter of the orbit through 3PN order, and PN and higher-order variations in the angle of nodes.  We also find, beginning at 3PN order, quadrupole and (spin)$^2$-induced variations in the eccentricity, semilatus rectum and inclination  [Eqs.\ (\ref{eq:dXdtheta}) and (\ref{eq:dXdtheta2})].   We obtain 3PN accurate expressions for the conserved energy, angular momentum component along the black-hole spin axis, and Carter constant, in terms of the average orbit elements [Eqs.\ (\ref{eq2:ELeCconserved})].  From the radiation-reaction sector (Sec.\ \ref{sec:reaction}), we obtain equations of evolution for the eccentricity and semilatus rectum that include the conventional ``Newtonian'' or quadrupole formula contributions (Peters-Mathews terms) at 2.5PN order, along with 3.5PN and 4.5PN terms, as well as 4PN spin-orbit terms [Eqs.\ (\ref{eq:dXdthetaRR})].   The latter terms change sign between prograde and retrograde orbits.   We also find spin-orbit induced, radiation-reaction variations at 4PN order in the inclination, pericenter and nodal angle.

Section \ref{sec:eccentricorbits} addresses an issue that arises with the osculating element formalism when the eccentricity tends to unity.  This limit is not well-behaved, in a manner similar to the well-known singular behavior of the eccentricity-pericenter pair of orbital elements in the limit of circular orbits.   We define a new eccentricity related to the old by PN corrections, and display the conserved quantities and the secular evolution equations in terms of this new variable.   

\begin{figure*}[t]
\begin{center}

\includegraphics[width=3.5in]{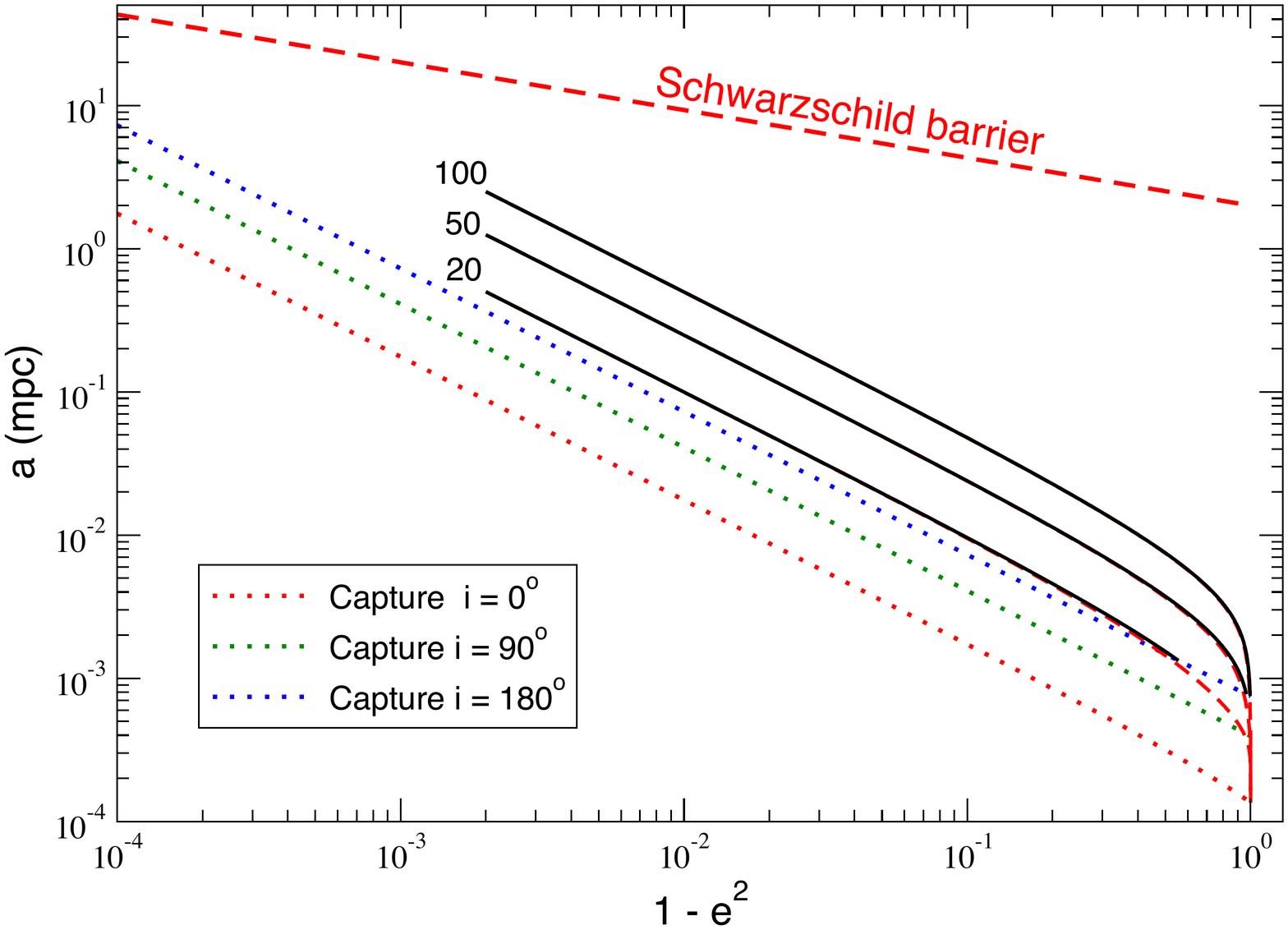}
\includegraphics[width=3.5in]{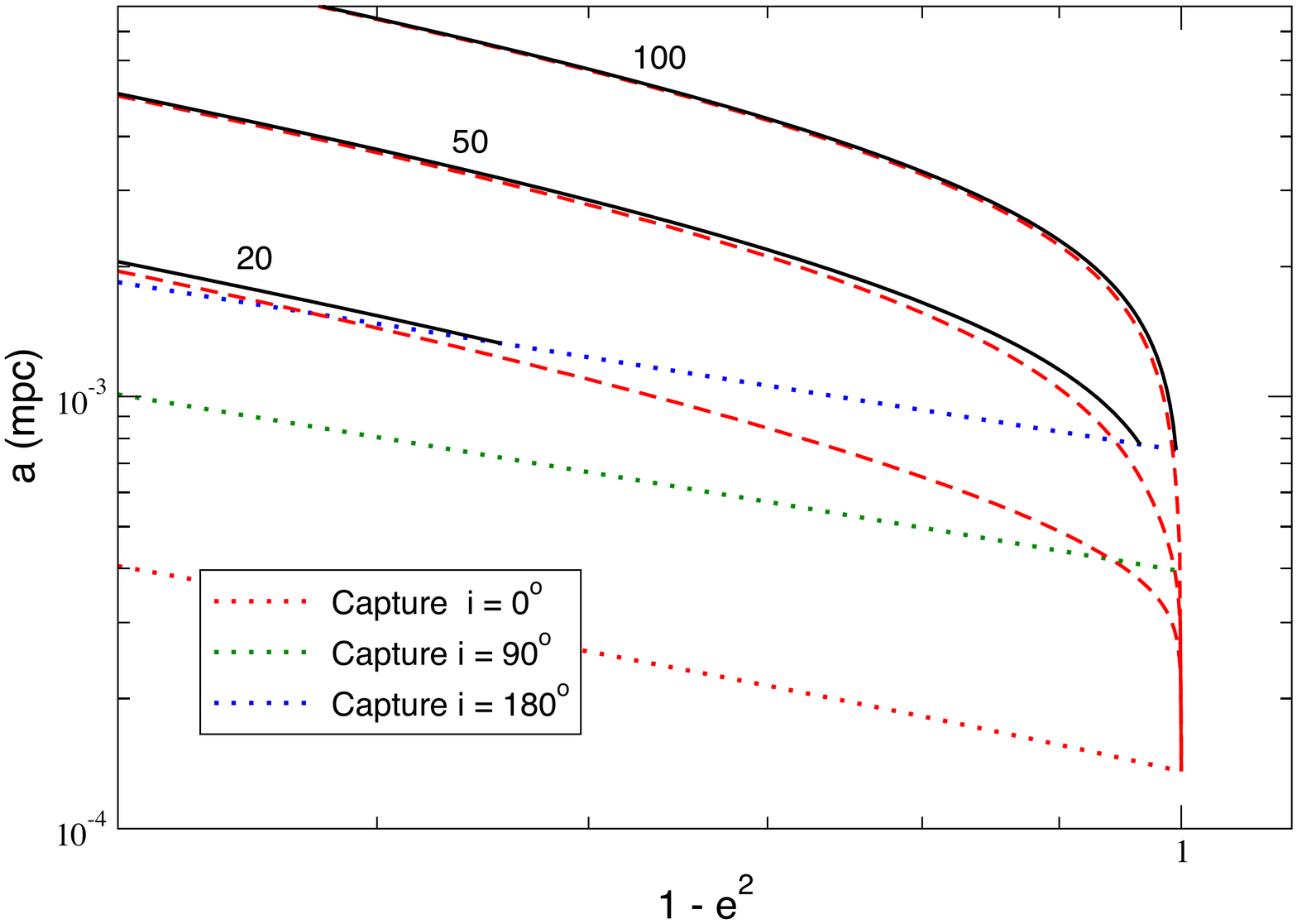}

\caption{\label{fig:merrittplots} Tracks of inspiral orbits with initial values of $p/(GM/c^2) = 100, \, 50$ and $20$, for a black hole of $M=10^6 M_\odot$ and $\chi=1$.  Dotted curves show the critical values for capture by the black hole from equatorial prograde (lower, red), polar (middle, green), and equatorial retrograde (top, blue) orbits.  Right:  late-time deviation of equatorial prograde (dashed, red) from equatorial prograde (solid, black) orbits, and their termination on the respective capture boundary. }
\end{center}
\end{figure*}

In Sec.\ \ref{sec:application}, we use these results to numerically evolve eccentric orbits from injection to plunge.   We recast the equations for gravitational radiation damping of the orbit in terms of dimensionless variables $\epsilon \equiv GM/c^2 p_i$, $x = p/p_i$, $e$, $\iota$, and $\theta$, where $p$, $e$ and $\iota$ are the semilatus rectum ($p_i$ is its initial value), eccentricity and inclination of the orbit, and $\theta$ is orbital phase.  These equations depend only on the spin parameter $\chi$, and thus the evolution in terms of $\theta$ depends only on $\epsilon$, $\chi$ and the initial conditions $e_i$ and $\iota_i$.  The black hole mass does not appear.  In Sec.\ \ref{sec:capture} we use our PN expression for the Carter constant along with results of \cite{2012CQGra..29u7001W} to establish a critical value of $p$ such that for $p < p_c$, the orbit will have no inner turning point and will plunge across the event horizon [Eq.\ (\ref{eq:condition})].  This will mark the termination of our numerical integrations.  Table \ref{tab:plunge} shows representative values of $p_c$ in units of $GM/c^2$ for various values of $\chi$ and $\iota$.

\begin{table}[b]
\caption{\label{tab:plunge} Critical semilatus rectum for capture}
\begin{ruledtabular}
\begin{tabular}{cccc}
Inclination $\iota$&\multicolumn{3}{c}{$p_c/(GM/c^2)$}\\
&$\chi=0$&$\chi=0.5$&$\chi=1$\\
\hline
$0$&9.04&6.09&2.71\\
$45^{\rm o}$&9.04&6.78&4.05\\
$90^{\rm o}$&9.04&8.77&7.84\\
$135^{\rm o}$&9.04&11.04&12.90\\
$180^{\rm o}$&9.04&12.03&14.98\\
\end{tabular}
\end{ruledtabular}
\end{table}

In Sec,\ \ref{sec:evolutionphase} we evolve the orbits with respect to phase, or number of orbits.
Figure \ref{fig:merrittplots} illustrates the results and places them  in an astrophysical context.  It is a  ``phase space'' diagram (in the terminology of those who study cluster dynamics) of energy vs angular momentum.  Here, semimajor axis $a = p/(1-e^2)$ is a proxy for energy and $1-e^2$ is a proxy for angular momentum.  The scale of the $a$-axis  assumes a $10^6 \, M_\odot$ black hole; we also assume that  $\chi =1$.  The dashed line (red in online version) shows schematically the Schwarzschild barrier, below which stellar orbits effectively decouple from resonant and non-resonant relaxation effects caused by perturbations from the other stars that drive random-walk behavior in eccentricity.   The precise location and slope of this barrier depends on the properties of the surrounding star cluster, and thus the curve shown is merely schematic.  Below this barrier, a stellar orbit is still subject to perturbations by other stars, of course, but we have chosen to ignore such perturbations.

 The three dotted curves show the critical values of $a$ vs. $1-e^2$ for plunge for equatorial prograde (lower, red), polar (middle, green), and equatorial retrograde (top, blue) orbits, respectively, as established by our capture criterion of Sec.\ \ref{sec:capture}.    These are curves of roughly constant $p$  or $a \propto (1-e^2)^{-1}$.   The three solid curves show the orbits in phase space of a body with $m=50 M_\odot$, with initial values of $p_i = 100, 50$ and $20 \,GM/c^2$, and with $e_i = 0.999$ for equatorial prograde (dashed, red) and retrograde orbits (solid, black).   Because $\eta = 5 \times 10^{-5}$, radiation reaction is a tiny effect initially, thus for all but the late phase of the evolution, $p$ and $e$ are approximately constant, thus $a$ for these orbits also varies inversely as $1-e^2$.   The difference between prograde and retrograde orbits is too small to be seen initially.   The right-hand panel of Figure \ref{fig:merrittplots} provides a blow-up of the late evolution of the orbits, showing the termination of each orbit when the condition for plunge is reached.   Retrograde orbits terminate at larger $a$ and larger $e$ than do prograde orbits.
We also obtain the number of orbits to plunge and the residual eccentricity as functions of $p_i/(GM/c^2)$ and $\iota$.  For initial values $e_i$ so close to unity, the results are independent of $e_i$.  While prograde orbits are highly circularized by the time of plunge because of their longer inspiral, retrograde orbits may be left with significant eccentricity at the time of plunge.  Table \ref{tab:eccentricity} shows residual values for $\iota = 180^{\rm o}$ (equatorial retrograde) orbits for a range of initial $p_i$.  

\begin{table}[b]
\caption{\label{tab:eccentricity} Residual eccentricity, $\iota = 180^{\rm o}$}
\begin{ruledtabular}
\begin{tabular}{cc}
Semilatus rectum&Residual Eccentricity\\
$p_i/(GM/c^2)$&$e$\\
\hline
20&0.48\\
40&0.18\\
60&0.10\\
80&0.07\\
100&0.05\\
\end{tabular}
\end{ruledtabular}
\end{table}

\begin{figure*}[t]
\begin{center}

\includegraphics[width=3.5in]{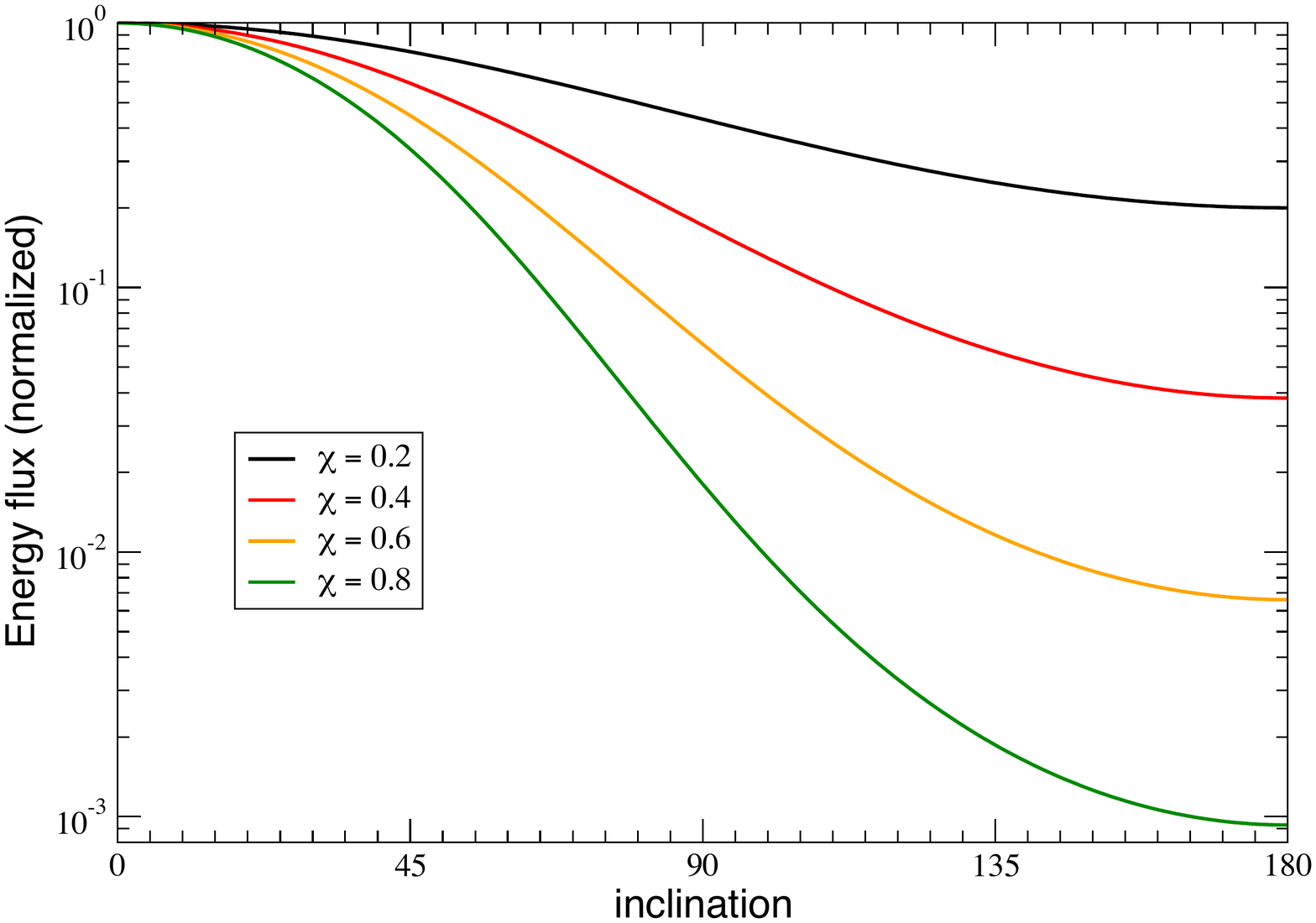}
\includegraphics[width=3.5in]{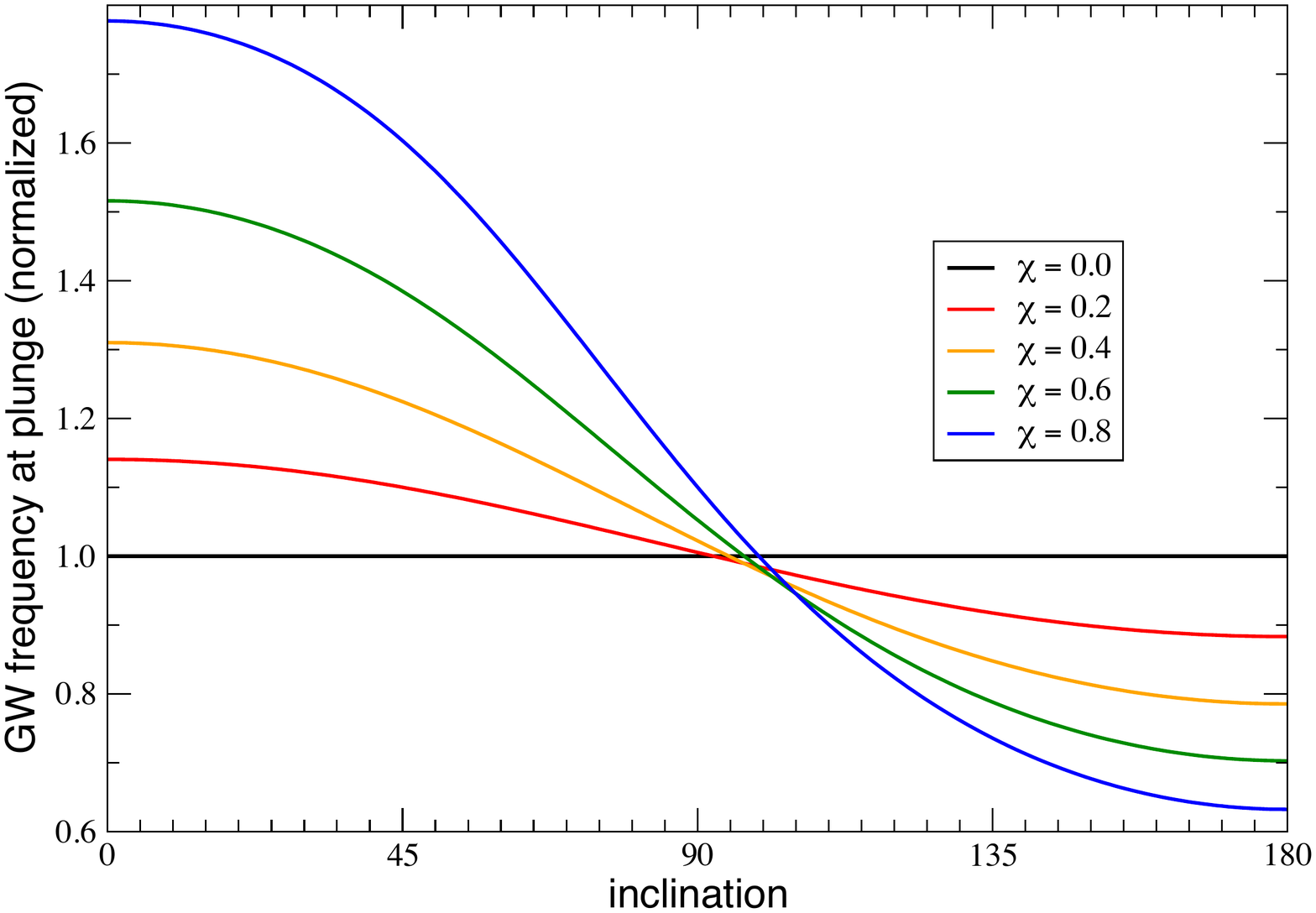}

\caption{\label{fig:fluxvsinc} Final-orbit gravitational-wave energy flux and frequency vs inclination for various Kerr spin parameters.  Fluxes are normalized to the value for $0^{\rm o}$ inclination, while frequencies are normalized to the value for $\chi = 0$. }  
\end{center}
\end{figure*}

Using an expression for the gravitational-wave flux accurate to 2PN order beyond the quadrupole approximation including spin-orbit contributions, we estimate the flux from the final orbit before plunge as a function of inclination.  Using a 2PN accurate expression for the orbital period $P$, we also estimate the gravitational-wave frequency at plunge.  The results are shown in Fig.\ \ref{fig:fluxvsinc}. 
We note that, for rapidly spinning black holes, there is a strong suppression of the energy flux for highly inclined and retrograde orbits compared to their prograde counterparts.  This is mostly due to the fact that inclined orbits plunge from greater distances than do equatorial prograde orbits, as can be seen in Fig.\ \ref{fig:merrittplots}.   For rapidly spinning black holes, the frequency can vary by as much as a factor of three between prograde and retrograde equatorial orbits, again because the former orbits survive to much closer to the spinning black hole than do the latter.

These  effects of inclination on the flux and frequency may have an impact on estimates of detectability of gravitational radiation from EMRIs by a space-based detector such as LISA.   Stars are likely to be injected into EMRI orbits with random inclinations.  Thus an estimate that uses only the gravitational-wave flux calculated by black hole perturbation theory for an equatorial prograde orbit in Kerr  before plunge, multiplied by a rate of injection of stars into EMRI orbits, may overestimate the total flux by failing to take the suppression induced by orbit inclination into account.    

\begin{figure*}[t]
\begin{center}

\includegraphics[width=5in]{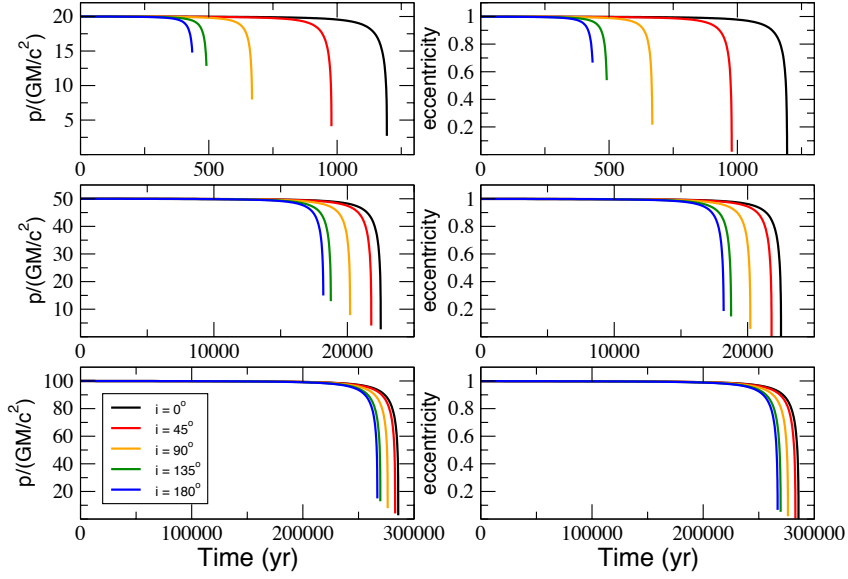}

\caption{\label{fig:pevstime} Evolution of semilatus rectum $p$ and eccentricity $e$ vs.\ time, for three initial values of $p$, for a $10^6 \, M_\odot$ black hole; in all cases, $e_i = 0.999$ and $\chi =1$. }
\end{center}
\end{figure*}

In Sec.\ \ref{sec:evolutiontime} we convert from $\theta$ to time using a 2PN accurate expression for orbital period $P$ and evolve the orbits with respect to time.  Fig.\ \ref{fig:pevstime} shows typical results, here for  $M = 10^6 M_\odot$, $\chi =1$, and $e_i = 0.999$.  Unlike the evolution with respect to orbital phase, the time to plunge is sensitive to $M$ and $e_i$ for a given initial $p_i$, because $P \propto M^{-1/2}  p_i^{3/2}  (1-e_i^2)^{-3/2}$.   In fact, the dependence is sufficiently simple that we were able to find an analytic approximation for the time to plunge, that agrees very well with the numerical results over many orders of magnitude of times, given by
\begin{eqnarray}
T_{\rm plunge} &=& \frac{1}{74.3 \, \eta} \left (\frac{GM}{c^3} \right ) G^\prime (e_i)  \epsilon^{-3.96} 
\nonumber \\
&& \quad \times \left ( 1 + 3\epsilon + 8 \epsilon^{3/2} \chi \cos \iota \right )^4 \,,
\label{eq:Tplungefit}
\end{eqnarray}
where $\epsilon = GM/c^2 p_i$ and 
\begin{equation}
G^\prime(e_i) =  \frac{3.35}{\sqrt{1-e_i^2}} -5 + 8 \sqrt{1-e_i^2}   \,.
\label{eq:G3}
\end{equation}
Equation (\ref{eq:Tplungefit}) can be incorporated into numerical codes that evolve stellar clusters around rotating black holes, providing a simple way to quickly determine the fate of a star that is injected into an orbit with an initial $p$, and $e$, and that suffers no further significant stellar perturbations.

The remainder of this paper provides the details underlying these results.  
Section \ref{sec:conclusions} makes concluding remarks.

\section{Post-Newtonian equations of motion for a body orbiting a supermassive rotating black hole} 
\label{sec:eom}

We study the motion of a body with a very small mass $m$ in the field of a supermassive rotating black hole of mass $M$ and spin $\bm S$.  At zeroth order in the mass of the orbiting body, the motion is described by the standard geodesic equation for the Kerr geometry.   These equations of motion are conservative, in that they admit conservation laws for energy $E$, angular momentum component $L_e$ in the direction of the spin axis $\bm e$,  and the so-called ``Carter constant'' $C$.    
   
We also include the effects of gravitational radiation reaction, which at lowest order are linear in the mass of the small body.   
This introduces a new parameter into the problem, conventionally denoted by the ``symmetric mass ratio'' $\eta$, defined by 
\begin{equation}
\eta \equiv \frac{Mm}{(M+m)^2} \sim \frac{m}{M} \, {\rm when} \, m \ll M \,.
\end{equation}
Formally this is inconsistent with our test-body conservative equations of motion, which assume $m = 0$.   However, in post-Newtonian theory, the equations of motion for non-spinning bodies are known to high PN orders for arbitrary masses, and in those equations  the coefficients are polynomials in powers of $\eta< 1/4$.  The problems that we wish to address with this work all involve stars or small black holes orbiting supermassive black holes so that $\eta \ll 1$.  Thus corrections due to finite masses in the conservative terms will be subdominant.   But the gravitational radiation reaction terms are proportional to $\eta$.  They are also dissipative, i.e. they violate the conservation of quantities associated with the orbit.   We will only be interested in the long-term dissipative effects of gravitational radiation at lowest order in $\eta$, and therefore we are justified in ignoring finite mass contributions to the conservative equations of motion. 

Throughout, we will work in harmonic coordinates, since those are the most suitable for discussing gravitational radiation and radiation reaction, and are the basis for post-Newtonian theory \cite{PW2014}.
Appendix \ref{app:Kerr} provides details for transforming the Kerr geometry from the conventional Boyer-Lindquist coordinates to harmonic coordinates, expanding the metric to 3PN order, and obtaining the test-body equations of motion and the conserved quantities $E$, $L_e$ and $C$ to 3PN order.  We make use of the fact that the black hole spin $\bm S$ and the Kerr parameter $a$ can be written as 
\begin{equation}
{\bm S} \equiv \frac{GM^2}{c} \chi {\bm e} \,, \quad a \equiv \frac{S}{M} = \frac{GM}{c} \chi \,,
\label{eq1:spindef}
\end{equation}
where $\chi$ is the dimensionless Kerr spin parameter bounded by $0 \le \chi \le 1$ and $\bm e$ is a unit vector parallel to the spin axis.   The geodesic equation to 3PN order is then given by
\begin{eqnarray}
\frac{d{\bm v}}{dt} &=& - \frac{GM{\bm n}}{r^2} 
\nonumber \\
&&
-  \frac{GM}{c^2 r^2} \left [  \left ( v^2 - 4\frac{GM}{r} \right ){\bm n} - 4 \dot{r} {\bm v} \right ]  
\nonumber \\
&&
+ \frac{GM^2}{c^3 r^3} \chi \left [6 {\bm e} \cdot ({\bm n} \times {\bm v}){\bm n} 
\right .
\nonumber \\
&& \qquad
\left .
+ 6\dot{r} ({\bm n} \times {\bm e}) -4({\bm v} \times {\bm e}) \right ]
\nonumber \\
&& 
-  \frac{G^2 M^2}{c^4 r^3} \biggl \{  \left [ \left( \frac{9GM}{r} - 2\dot{r}^2 \right ) {\bm n} + 2 \dot{r} {\bm v} \right ]
\nonumber \\
&&
\qquad 
-  \frac{3}{2} \frac{GM}{r} \chi^2 \left [ 5{\bm n}({\bm e} \cdot {\bm n})^2 - 2{\bm e}({\bm e} \cdot {\bm n}) - {\bm n} \right ]  \biggr \}
\nonumber \\
&&
- \frac{G^2 M^2}{c^5 r^3} \chi \biggl \{\frac{GM}{r} \biggl ( 20 {\bm e} \cdot ({\bm n} \times {\bm v}){\bm n} 
\nonumber \\
&&
\qquad
+ 16\dot{r} ({\bm n} \times {\bm e})
-12({\bm v} \times {\bm e}) \biggr )
\nonumber \\
&&
\qquad
+6\dot{r} {\bm e} \cdot ({\bm n} \times {\bm v}) {\bm v} \biggr \}
\nonumber \\
&& 
+  \frac{G^3 M^3}{c^6 r^4} \biggl \{ 
\left [ \left( \frac{16GM}{r} - \dot{r}^2 \right ) {\bm n}
+ 4 \dot{r} {\bm v} \right ]
\nonumber \\
&&
\qquad 
+ \chi^2 \biggl [ \frac{3}{2} \left ( 5{\bm n}({\bm e} \cdot {\bm n})^2 - 2{\bm e}({\bm e} \cdot {\bm n}) - {\bm n} \right ) 
\nonumber \\
&&
\qquad \qquad 
\times  \left (v^2 - \frac{4GM}{r} \right )
\nonumber \\
&&
\qquad
- 6 {\bm v}  \left ( 5\dot{r}({\bm e} \cdot {\bm n})^2 - 2({\bm v} \cdot {\bm e})({\bm e} \cdot {\bm n}) - \dot{r} \right ) 
\nonumber \\
&&
\qquad
+ \frac{2 GM }{r}  \left ( {\bm n} -6{\bm n}({\bm e} \cdot {\bm n})^2+ ({\bm e} \cdot {\bm n}) {\bm e}  \right ) \biggr ] 
\biggr \}
\,.
\nonumber \\
\label{eq:eom2}
\end{eqnarray}
 ``Monopole'', or pure mass PN terms occur at the usual 1PN ($c^{-2}$), 2PN ($c^{-4}$) and 3PN ($c^{-6}$) orders.  Spin-orbit terms linear in $\chi$ occur at 1.5PN ($c^{-3}$) and 2.5PN ($c^{-5}$) orders, shifted as usual by a half PN order relative to their counterparts for normal slowly rotating bodies.  The 2PN term quadratic in $\chi$ is the leading-order contribution of the black hole's ``Newtonian'' quadrupole moment, while the 3PN terms quadratic in $\chi$ are a combination of (spin)$^2$ terms and of  ``cross-terms'' between  the quadrupole and monopole terms.  

We now supplement these equations with the contributions from gravitational radiation reaction.
These include 2.5PN terms ($c^{-5}$), 3.5PN terms ($c^{-7}$) \cite{2002PhRvD..65j4008P}, a 4PN spin-orbit contribution ($c^{-8}$) \cite{2005PhRvD..71h4027W} and 4.5PN terms ($c^{-9}$) \cite{1997PhRvD..55.6030G}.  
They are given, in harmonic coordinates, by
 \begin{eqnarray}
{\frac{d{\bm v}}{dt}}_{\rm RR} &=& \frac{8}{5} \eta  \frac{G^2 M^2}{c^5 r^3} \left [ \left (3 v^2 + \frac{17}{3} \frac{GM}{r} \right )\dot{r} {\bm n}  
\right .
\nonumber \\
&& \quad \quad
\left .
- \left ( v^2 + 3 \frac{GM}{r} \right ) {\bm v} \right ]
\nonumber \\
&&
- \frac{8}{5} \eta  \frac{G^2 M^2}{c^7 r^3} \biggl [ \biggl (
\frac{183}{28} v^4+\frac{519}{42} v^2 \frac{GM}{r} 
\nonumber \\
&&
\quad \quad
- \frac{285}{4} v^2 \dot{r}^2 +\frac{147}{4} \dot{r}^2 \frac{GM}{r} 
\nonumber \\
&&
\quad \quad
+ 70 \dot{r}^4 + \frac{989}{14} \frac{G^2M^2}{r^2} \biggr ) \dot{r} {\bm n} 
\nonumber \\
&&
\qquad
- \biggl ( \frac{313}{28} v^4-\frac{205}{42} v^2 \frac{GM}{r} - \frac{339}{4} v^2 \dot{r}^2
\nonumber \\
&&
\qquad 
+\frac{205}{12} \dot{r}^2 \frac{GM}{r} + 75 \dot{r}^4 + \frac{1325}{42} \frac{G^2M^2}{r^2}
\biggr ) {\bm v} \biggr ]
\nonumber \\
&&
-\frac{1}{5} \eta  \frac{G^3 M^3}{c^8 r^4} \chi \biggl \{ 
{\bm e} \cdot ({\bm n} \times {\bm v}) 
\nonumber \\
&&
\qquad \quad
\times \left (A_{\rm 4SO}\, \dot{r} {\bm n} + B_{\rm 4SO}\, {\bm v} \right )
\nonumber \\
&&
\qquad
-\frac{2}{3} ({\bm v} \times {\bm e} ) \dot{r} C_{\rm 4SO} + \frac{1}{2}  ({\bm n} \times {\bm e} ) D_{\rm 4SO} \biggr \}
\nonumber \\
&&
+  \frac{8}{5} \eta \frac{G^2 M^2}{c^9 r^3} \biggl [ E_{\rm 4.5} \dot{r} {\bm n} - F_{\rm 4.5} {\bm v} \biggr ]\,.
\label{eq:eomRR}
\end{eqnarray}
The expressions for $A_{\rm 4SO}$, $B_{\rm 4SO}$, $C_{\rm 4SO}$, $D_{\rm 4SO}$, $E_{\rm 4.5}$ and $F_{\rm 4.5}$ are given in Appendix \ref{app:Kerr}.  For simplicity we do not include the non-local 4PN ``tail'' term.

In studying the evolution of the system under radiation reaction, we cannot simply ignore the conservative PN test-body terms displayed in Eq.\ (\ref{eq:eom2}), and treat the problem as if it were Newtonian gravity plus radiation reaction.  That is because, in solving for the secular evolution of the orbital elements, there will be ``cross term'' interactions between, say the 1PN conservative perturbations or the 1.5PN spin-orbit terms and the 2.5PN radiation-reaction terms, that could produce dissipative contributions at 3.5PN order or 4PN spin-orbit order, respectively.  These interactions {\em must} be taken into account to get the complete higher-order evolution.  

\section{Lagrange planetary equations for the osculating orbit elements}
\label{sec:lagrange}

\subsection{Basic equations}
\label{sec:basiceq}

We begin with a brief review of standard orbital perturbation theory, used to compute deviations from Keplerian two-body motion induced by perturbing forces, described by the equation of motion
\begin{equation}
\frac{d^2 {\bm x}}{dt^2} = - \frac{GM}{r^3}{\bm x} + \delta {\bm a} \,,
\label{eq2:eom}
\end{equation}
where $\delta {\bm a}$ is a perturbing acceleration.  
For a general orbit described by $\bm x$ and $\bm v = d{\bm x}/dt$, we define the ``osculating'' Keplerian orbit using a set of orbit elements: the semi-latus rectum $p$, eccentricity $e$, inclination $\iota$, nodal angle $\Omega$ and pericenter angle $\omega$, defined by the following set of equations:
\begin{align}
{\bm x} &\equiv r {\bm n} \,,
\nonumber \\
r &\equiv p/(1+e \cos f) \,,
\nonumber \\
{\bm n} &\equiv \left ( \cos \Omega \cos \phi- \cos \iota \sin \Omega \sin \phi \right ) {\bm e}_X 
\nonumber \\
& \quad
 + \left ( \sin \Omega \cos \phi + \cos \iota \cos \Omega \sin \phi \right ){\bm e}_Y
\nonumber \\
& \quad
+ \sin \iota \sin \phi {\bm e}_Z \,,
\nonumber \\
{\bm \lambda} &\equiv \partial {\bm n}/\partial \phi \,, \quad \hat{\bm h}={\bm n} \times {\bm \lambda} \,,
\nonumber \\
{\bm h} &\equiv {\bm x} \times {\bm v} \equiv \sqrt{GMp} \, \bm{\hat{h}} \,,
\label{eq2:keplerorbit}
\end{align}
where  $f \equiv \phi - \omega$ is the  {\em true anomaly}, $\phi$ is the orbital phase measured from the ascending node, and 
 ${\bm e}_A$ are chosen reference basis vectors.   From the given definitions, we see that ${\bm v} = \dot{r} {\bm n} + (h/r) {\bm \lambda}$ and $\dot{r} = (he/p) \sin f$.   We also define the alternative orbit elements:
 \begin{eqnarray}
\alpha &=& e \cos \omega \,,
\nonumber \\
\beta &=& e \sin \omega \,, 
\label{eq2:alphabeta}
\end{eqnarray}
so that now
\begin{equation}
r = p/(1+\alpha \cos \phi + \beta \sin \phi) \,.
\label{eq2:rdef}
\end{equation}

One then defines the radial $\cal R$, cross-track $\cal S$ and out-of-plane $\cal W$ components of the perturbing acceleration $\delta {\bm a}$, defined respectively by
${\cal R} \equiv {\bm n} \cdot \delta {\bm a}$,
 ${\cal S} \equiv {\bm \lambda} \cdot \delta {\bm a}$ and
 ${\cal W} \equiv \bm{\hat{h}} \cdot \delta {\bm a}$,
and writes down the ``Lagrange planetary equations'' for the evolution of the orbit elements,
\begin{eqnarray}
\frac{dp}{dt} &=& 2 \sqrt{\frac{p}{GM}} \,r {\cal S}\,,
\nonumber \\
\frac{d\alpha}{dt} &=& \sqrt{\frac{p}{GM}} \left [ {\cal R}  \sin \phi   +  {\cal S} (\alpha + \cos \phi) \left ( 1+ \frac{r}{p} \right ) 
\right .
\nonumber \\
&&
\left . 
\quad 
+ {\cal W} \frac{r}{p} \beta \cot \iota  \sin \phi \right ]\,,
\nonumber \\
\frac{d\beta}{dt} &=& \sqrt{\frac{p}{GM}} \left [ -{\cal R}  \cos \phi   +  {\cal S} (\beta + \sin \phi) \left ( 1+ \frac{r}{p} \right ) 
\right .
\nonumber \\
&&
\left . 
\quad
- {\cal W} \frac{r}{p} \alpha \cot \iota  \sin \phi \right ]\,,
\nonumber \\
\frac{d\iota}{dt} &=& \sqrt{\frac{p}{GM}} {\cal W} \left (\frac{r}{p} \right )\cos \phi \,,
\nonumber \\
\frac{d\Omega}{dt} &=& \sqrt{\frac{p}{GM}} {\cal W} \left (\frac{r}{p} \right ) \frac{\sin \phi}{\sin \iota } \,,
\nonumber \\
\frac{d\phi}{dt} &=& \frac{h}{r^2} -  \cos \iota \frac{d\Omega}{dt}  \,,
\label{eq2:Lagrange}
\end{eqnarray}
(see \cite{PW2014} for further discussion.)  The final equation is not really an orbit element equation, but serves to close the system; in some formulations the sixth equation determines the 	``time of pericenter passage''.
These equations are {\em exact}: they are simply a reformulation of Eq.\ (\ref{eq2:eom}) in terms of new variables.  However, they are particularly useful when the perturbations are small, so that solutions can be obtained by a process of iteration.  At lowest order (no perturbations), the elements $p$, $\alpha$, $\beta$, $\Omega$ and $\iota$ are constant, and $d\phi/dt = \sqrt{GMp}/r^2$; those solutions can be plugged into the right-hand-side and the equations integrated to find corrections, and so on.  The corrections to the orbit elements tend to be of two classes: {\em periodic} corrections, which vary on an orbital timescale, and {\em secular} corrections, which vary on a longer timescale, depending upon the nature of the perturbations.   There are many approaches to separating secular from periodic effects (see \cite{1961mcm..book.....B}, for example), but we will adopt a specific approach, known as ``multiple-scale analysis'', that is used in many problems in physics \cite{1978amms.book.....B}.   One advantage of this method is that it can be carried out systematically to higher orders in perturbation theory. 

We have formulated the Lagrange planetary equations using $\alpha$ and $\beta$ instead of $e$ and $\omega$ because of the well-known problem that the latter formulation is singular in the limit $e \to 0$ (the pericenter angle $\omega$ is formally undefined in this limit).    At 1PN order, for example, the osculating orbit that corresponds to a true circular orbit ($r=$ constant) is one in which $e \approx 3GM/c^2 p$, the pericenter advances at the same rate as the orbital motion, $\omega \propto \phi$, and the true anomaly is perpetually fixed at apocenter, $f=\pi$ \cite{1990PhRvD..42.1123L}.  The fact that $\omega$ no longer experiences a small precession illustrates the singular nature of this limit.  By contrast, the $\alpha-\beta$ formulation is completely regular in the limit $e = (\alpha^2 +\beta^2)^{1/2} =0$.  An alternative approach involves defining {\em two} eccentricities, one associated with the radial motion and one with the angular motion \cite{1985AIHS...43..107D,1986AIHS...44..263D}.   However, we will stick with the $\alpha,\, \beta$ orbit elements.

We now use the sixth planetary equation to express the remaining five equations as differential equations in $\phi$, in the general form
\begin{equation}
\frac{d X_\alpha (\phi)}{d\phi} = \epsilon Q_\alpha (X_\beta(\phi), \phi) \,,
\label{eq2:dXdf}
\end{equation}
where $\alpha, \, \beta$ label the orbit element, $\epsilon$ is a small parameter that characterizes the perturbation, and
\begin{equation}
Q_\alpha (X_\beta(\phi), \phi) = \frac{(r^2/h) Q^{(t)}_\alpha}{1-(r^2/h) \cos \iota Q^{(t)}_\Omega } \,,
\label{eq:Qalpha}
\end{equation}
where
the $Q^{(t)}_\alpha$, denote the right-hand-sides of Eqs.\ (\ref{eq2:Lagrange}).  Note that, because we are working to higher PN orders, we {\em must}, at least in principle, include the additional factor involving $Q^{(t)}_\Omega$.

\subsection{Two-timescale analysis of the planetary equations}
\label{sec:twoscale}

We anticipate that the solutions for the $X_\alpha$ will have pieces that vary on a ``short'' orbital timescale, corresponding to the periodic functions of $\phi$, but may also have pieces that vary on a long timescale, of order $1/\epsilon$ times the short timescale.   In a two-timescale analysis \cite{1978amms.book.....B,1990PhRvD..42.1123L,2004PhRvD..69j4021M,2008PhRvD..78f4028H}, one treats these two times formally as independent variables, and solves the ordinary differential equations as if they were partial differential equations for the two variables.
We define the long timescale variable
\begin{equation}
\theta \equiv \epsilon \phi \,,
\label{eq2:thetadef}
\end{equation}
and write the derivative with respect to $\phi$ as
\begin{equation}
\frac{d}{d\phi} \equiv \epsilon \frac{\partial}{\partial \theta} + \frac{\partial}{\partial \phi} \,.
\label{eq2:ddphi}
\end{equation}
We make an {\em ansatz} for the solution for $X_\alpha (\theta, \phi)$:
\begin{equation}
X_\alpha (\theta, \phi) \equiv \tilde{X}_\alpha (\theta) + \epsilon Y_\alpha (\tilde{X}_\beta (\theta), \phi) \,.
\label{eq2:ansatz}
\end{equation}
On short timescales, this would represent the standard solution: an initial value of the variable $\tilde{X}_\alpha$, plus a solution that depends on the initial conditions.  But on longer timescales, these ``initial'' values are allowed to vary.   

This split of $X_\alpha$ is not unique.  However, if the explicit $\phi$ dependence in $Q_\alpha$ is periodic, a natural choice to define the split is to assume that 
\begin{equation}
\tilde{X}_\alpha (\theta) = \langle X_\alpha (\theta, \phi) \rangle \,, \quad \langle Y_\alpha (\tilde{X}_\beta (\theta), \phi) \rangle = 0 \,,
\label{eq2:split}
\end{equation}
where the ``average'' $\langle \dots \rangle$ is defined by 
\begin{equation}
\langle A \rangle \equiv \frac{1}{2\pi} \int_0^{2\pi} A(\theta,\phi) d\phi \,,
\label{eq2:averagedef}
\end{equation}
holding $\theta$ fixed.  
For any function $A(\theta,\phi)$ we define the ``average-free'' part as
\begin{equation}
 {\cal AF}(A) \equiv  A(\theta,\phi) - \langle A \rangle   \,.
 \label{eq2:averagefreedef}
\end{equation}
We now substitute Eq.\ (\ref{eq2:thetadef}) into (\ref{eq2:dXdf}), and divide by the parameter $\epsilon$, to obtain
\begin{equation}
\frac{d \tilde{X}_\alpha (\theta)}{d\theta} + \frac{\partial Y_\alpha}{\partial \phi} + \epsilon \frac{\partial Y_\alpha}{\partial \tilde{X}_\gamma}
\frac{d\tilde{X}_\gamma}{d\theta} = Q_\alpha (\tilde{X}_\beta + \epsilon Y_\beta, \phi) \,,
\label{eq2:split2}
\end{equation}
where we sum over the range of $\gamma$.
Taking the average and average-free parts of this equation, we obtain
\begin{subequations}
\begin{align}
\frac{d\tilde{X}_\alpha}{d\theta} &= \langle Q_\alpha (\tilde{X}_\beta + \epsilon Y_\beta, \phi) \rangle \,,
\label{eq2:aveq}\\
\frac{\partial Y_\alpha}{\partial \phi} &= {\cal AF} \left (Q_\alpha (\tilde{X}_\beta + \epsilon Y_\beta, \phi) \right )  - \epsilon \frac{\partial Y_\alpha}{\partial \tilde{X}_\gamma} \frac{d\tilde{X}_\gamma}{d\theta} \,.
\label{eq2:avfreeeq}
\end{align}
\label{eq2:maineq}
\end{subequations}
These equations can then be iterated in a straightforward way.  At zeroth order, Eq.\ (\ref{eq2:aveq}) yields $d\tilde{X}_\alpha/d\theta = \langle Q^0_\alpha  \rangle$ where $Q^0_\alpha \equiv Q_\alpha (\tilde{X}_\beta, \phi)$, which is the conventional result whereby one averages the perturbation holding the orbit elements fixed.  We write the expansion $Y_\alpha  \equiv Y^{(0)}_\alpha + \epsilon Y^{(1)}_\alpha + \epsilon^2 Y^{(2)}_\alpha + \dots$.  We then integrate Eq.\ (\ref{eq2:avfreeeq}) holding $\theta$ fixed to obtain $Y^0_\alpha$.  Appendix \ref{app:twoscale} provides details on integrating this equation with boundary conditions chosen to ensure that the answer is average-free.  The iteration continues until one obtains all contributions to  $d\tilde{X}_\alpha/d\theta$ compatible with the order in $\epsilon$ to which $Q_\alpha$ is known.   The final solution including periodic terms is given by Eq.\ (\ref{eq2:ansatz}), with the secular evolution of the $\tilde{X}_\alpha$ given by solutions of Eqs.\ (\ref{eq2:aveq}).  From these solutions one can reconstruct the instantaneous orbit using Eqs.\ (\ref{eq2:keplerorbit}).  

If one is interested only in the secular evolutions to a chosen order, then one can Taylor expand $Q_\alpha$ in Eqs.\ (\ref{eq2:maineq}) in powers of $\epsilon$ and carry out the iterations explicitly.  Through order $\epsilon^2$ beyond the lowest order term, the result is 
\begin{eqnarray}
\frac{d\tilde{X}_\alpha}{d\phi} &=& \epsilon \langle Q_\alpha^{(0)} \rangle + \epsilon^2 \biggl [ \langle Q_{\alpha,\beta}^{(0)} \int_0^\phi Q_{\beta}^{(0)} d\phi' \rangle 
\nonumber \\
&&
+  \langle Q_{\alpha,\beta}^{(0)} \rangle \langle \phi\,Q_\beta^{(0)} \rangle
-  \langle \phi\,Q_{\alpha,\beta}^{(0)} \rangle \langle Q_\beta^{(0)} \rangle
\nonumber \\
&&  
- \pi \langle Q_{\alpha,\beta}^{(0)}  \rangle \langle Q_\beta^{(0)} \rangle  \biggr ] + O(\epsilon^3)\,,
\label{eq2:dXdtfinal}
\end{eqnarray}
where the subscript $,\beta$ denotes $\partial/\partial \tilde{X}_\beta$, and where we have converted from $\theta$ back to $\phi$.   Since we are working to 3PN order in the conservative dynamics and to 4.5PN order in the radiation-reaction sector, we will need to include the $O(\epsilon^3)$ contributions; Appendix \ref{app:twoscale} provides details.

\begin{widetext}

\subsection{Secular conservative dynamics}
\label{sec:conservative}

We first carry out this procedure for the conservative part of the equations of motion.  At zeroth order in $\epsilon$, we obtain
\begin{eqnarray}
\frac{d\tp}{d\theta} &=& \frac{d\ti}{d\theta} = \frac{d\tOm}{d\theta} = 0 \,,
\nonumber \\
\frac{d\tal}{d\theta} &=& -\frac{3GM}{c^2 \tp} \tbe \,,
\nonumber \\
\frac{d\tbe}{d\theta} &=& \frac{3GM}{c^2 \tp} \tal \,,
\label{eq2:zerothorderav}
\end{eqnarray}
from which we obtain $d\te/d\theta = 0$ and $d\tom/d\theta = 3GM/c^2 \tp$, the standard 1PN results for Schwarzschild.  For the average-free part, we obtain, including the 1.5PN order spin-orbit contribution,

\begin{eqnarray}
Y^0_p &=& -8\frac{GM}{c^2} (\tal \cos \phi + \tbe \sin \phi ) + 4 \tp \left ( \frac{GM}{c^2 \tp} \right )^{3/2}
\chi \cos \ti (\tal \cos \phi + \tbe \sin \phi )
\,,
\nonumber \\
Y^0_\alpha &=& -\frac{GM}{2c^2 \tp} \left [  5\tbe \sin 2\phi +5\tal \cos 2\phi  +2 (3+7\tal^2-\tbe^2) \cos \phi  + 16 \tal \tbe \sin \phi \right ] 
\nonumber \\
&& \qquad
+ \frac{1}{2} \left ( \frac{GM}{c^2 \tp} \right )^{3/2} \chi \cos \ti  \left [ \tbe^2 \cos 3\phi - \tal \tbe \sin 3\phi
-2 \tbe \sin 2\phi - (4 - 4 \tal^2 + 9\tbe^2 ) \cos \phi +11 \tal \tbe \sin \phi  \right ]
\,,
\nonumber \\
Y^0_\beta &=& -\frac{GM}{2c^2 \tp} \left [  5\tal \sin 2\phi - 5\tbe \cos 2\phi +2 (3+7\tbe^2-\tal^2) \sin \phi 
+ 16 \tal \tbe \cos \phi \right ] 
\nonumber \\
&& \qquad
+ \frac{1}{2} \left ( \frac{GM}{c^2 \tp} \right )^{3/2} \chi \cos \ti  \left [ \tal^2 \sin 3\phi - \tal \tbe \cos 3\phi  + 2 \tal \sin 2\phi - (4 - 4 \tbe^2 + 7\tal^2 ) \sin \phi +13 \tal \tbe \cos \phi  \right ]
\,,
\nonumber \\
Y^0_\iota &=& 
- \frac{1}{2} \left ( \frac{GM}{c^2 \tp} \right )^{3/2} \chi \sin \ti  \left [ \tal \cos 3\phi + \tbe \sin 3\phi  + 2 \cos 2\phi  +3 \tal  \cos \phi  + \tbe \sin \phi   \right ] \,,
\nonumber \\
Y^0_\Omega &=& -  \left ( \frac{GM}{c^2 \tp} \right )^{3/2} \chi
\left [ \sin 2 \phi (1+ \tal \cos \phi + \tbe \sin \phi ) -2 \tal \sin \phi + 2 \tbe \cos \phi   \right ] \,.
\label{eq2:zerothorderavfree}
\end{eqnarray}
The solution for the orbital separation as a function of $\phi$ is given by
\begin{equation}
r^{-1} = \frac{1 + (\tal + Y^0_\alpha) \cos \phi +  (\tbe + Y^0_\beta) \sin \phi}{\tp + Y^0_p} \,,
\end{equation}
In the limit $\tal = \tbe = 0$, this yields
\begin{eqnarray}
r^{-1} &=& \frac{1  - 3(GM/c^2 \tp) -2 (GM/c^2 \tp)^{3/2} \chi \cos \ti }{\tp} \,,
\end{eqnarray}
corresponding to a circular orbit, without any of the singular behavior associated with $\tom$.  On the other hand, in the limit $\te \to 1$, we obtain
\begin{equation}
r^{-1} = \frac{1+\cos f}{\tp} \left [ 1 + \frac{11}{2} \frac{GM}{c^2 \tp} - \left (\frac{GM}{c^2 \tp} \right )^{3/2} \chi \cos \ti \left ( 6 - 2 \cos f - \sin f \sin 2\phi \right ) \right ]  - \frac{15}{2} \frac{GM}{c^2 \tp^2} \,.
\label{eq:rinve1}
\end{equation}
where $f = \phi - \tom$.   Because of the presence of the final 1PN term, $r^{-1}$ does not vanish in the obvious manner at $f = \pi$, as might be expected for an unbound orbit, but rather at $f=\pi + (15GM/c^2 \tp)^{1/2}$.   We will address this and other unusual behavior of the solutions as $\te \to 1$ in Sec.\ \ref{sec:eccentricorbits}. 

Carrying out the iterations to an order consistent with the 3PN order of the equations of motion, we obtain finally
\begin{subequations}
\begin{eqnarray}
\frac{d\tp}{d\theta} &=&  - 6 \tp \left ( \frac{GM}{c^2 \tp} \right )^3 \tal \tbe \chi^2 \sin^2  \ti \,,
\label{eq:dpdtheta}
 \\
\frac{d\tal}{d\theta} &=& -\frac{3GM}{c^2 \tp} \tbe 
+ 6 \left (\frac{GM}{c^2 \tp} \right )^{3/2}  \tbe \,\chi \, \cos \ti 
+\frac{3}{4} \left ( \frac{GM}{c^2 \tp} \right )^2 \tbe \left [ 10 - \tal^2 - \tbe^2 -\chi^2 (5 \cos^2 \ti -1) \right ]
\nonumber \\
&& - 3 \left ( \frac{GM}{c^2 \tp} \right )^{5/2} \tbe (8 - 3\tal^2 - 3\tbe^2) \chi \cos \ti
\nonumber \\
&&
- \frac{3}{2}  \left ( \frac{GM}{c^2 \tp} \right )^{3} \tbe \left [ (29 + 34\tal^2 + 34\tbe^2) +\frac{1}{4} \chi^2 \left ( 7 + 2\tal^2 - 18\tbe^2 - [23 - 62 \tal^2 - 90\tbe^2 ]  \cos^2 \ti \right ) \right ]
\,,
\label{eq:daldtheta}
 \\
\frac{d\tbe}{d\theta} &=& \frac{3GM}{c^2 \tp} \tal 
- 6 \left (\frac{GM}{c^2 \tp} \right )^{3/2} \tal \, \chi \, \cos \ti  
-\frac{3}{4} \left ( \frac{GM}{c^2 \tp} \right )^2 \tal \left [ 10 - \tal^2 - \tbe^2 -\chi^2 (5 \cos^2 \ti -1) \right ]
\nonumber \\
&& + 3 \left ( \frac{GM}{c^2 \tp} \right )^{5/2} \tal (8 - 3\tal^2 - 3\tbe^2) \chi \cos \ti
\nonumber \\
&&
+ \frac{3}{2}  \left ( \frac{GM}{c^2 \tp} \right )^{3} \tal \left [ (29 + 34\tal^2 + 34\tbe^2) -\frac{1}{4} \chi^2 \left ( 3 + 6\tal^2 + 26\tbe^2 + [13 - 70 \tal^2 - 98\tbe^2 ]  \cos^2 \ti \right ) \right ]
\,,
\label{eq:dbedtheta}
 \\
 \frac{d\ti}{d\theta} &=& -3 \left ( \frac{GM}{c^2 \tp} \right )^{3} \tal \tbe \chi^2 \sin \ti \cos \ti \,,
\label{eq:didtheta}
\\
 \frac{d\tOm}{d\theta} &=&  2 \left (\frac{GM}{c^2 \tp} \right )^{3/2} \chi -\frac{3}{2} \left ( \frac{GM}{c^2 \tp} \right )^2 \chi^2 \cos \ti - 3 \left ( \frac{GM}{c^2 \tp} \right )^{5/2} \chi (4-\tal^2 - \tbe^2)
\nonumber \\
&&
+ \frac{3}{2} \left ( \frac{GM}{c^2 \tp} \right )^{3} (8 - 7\tal^2-9\tbe^2) \chi^2 \cos \ti  \,.
\label{eq:dOmdtheta}
\end{eqnarray}
\label{eq:dXdtheta}
\end{subequations}
We also obtain, but do not display, the average-free, periodic contributions $Y_\alpha$ through 3PN order.  From the equations for $\alpha$ and $\beta$ we obtain
\begin{subequations}
\begin{eqnarray}
\frac{d\tom}{d\theta} &=& 3 \frac{GM}{c^2 \tp} - 6 \left ( \frac{GM}{c^2 \tp} \right )^{3/2} \chi \cos \ti
- \frac{3}{4} \left ( \frac{GM}{c^2 \tp} \right )^{2} \left [ 10 - \te^2 + \chi^2 \left ( 1-5\cos^2 \ti \right ) \right ]
\nonumber \\
&&
+ 3 \left ( \frac{GM}{c^2 \tp} \right )^{5/2} \left (8 - 3 \te^2 \right ) \chi \cos \ti 
+ \frac{3}{8} \left ( \frac{GM}{c^2 \tp} \right )^{3} \biggl \{ 4 \left (29 + 34 \te^2 \right ) - \chi^2 \left [ 4(4 -17 \te^2 ) - 2 (9 - 40 \te^2) \sin^2 \ti 
\right .
\nonumber \\ 
&&
\left .
\qquad + \left (4 \te^2 + 5(1-2 \te^2) \sin^2 \ti \right ) \cos (2\tom) \right ] \biggr \} \,,
\label{eq:domdtheta2} \\
\frac{d\te}{d\theta} &=& -\frac{3}{4} \left ( \frac{GM}{c^2 \tp} \right )^{3} \te (5 + 4\te^2) \chi^2 \sin^2 \ti \sin \tom \cos \tom \,.
\label{eq:dedtheta2}
\end{eqnarray}
\label{eq:dXdtheta2}
\end{subequations}

Substituting the definitions (\ref{eq2:keplerorbit}) of the osculating orbits into the expressions (\ref{app:E3pn}), (\ref{app:Le3pn}), and (\ref{app:C3pn}) for $E$, $L_e$ and $C$, we obtain $E = -GM(1-\te^2)/2\tp + \dots$, $L_e = (GM\tp)^{1/2} \cos \ti + \dots$, and $C = (GM\tp) + \dots$, where the PN corrections depend in general on the phase $\phi$.  However, now substituting the solutions for each orbit element $X_\alpha = \tilde{X}_\alpha + Y_\alpha ( \tilde{X}_\beta ,\phi)$, with the $Y_\alpha$ computed through 3PN order, we find that all $\phi$ dependence cancels, and the conserved quantities are given purely in terms of the  $\tilde{X}_\alpha$:
\begin{subequations}
\begin{eqnarray}
c^{-2} E &=& -\frac{1}{2} \frac{GM(1-\te^2)}{c^2 \tp} + \frac{1}{8} \left (\frac{GM}{c^2 \tp} \right )^2 \left ( 19+38 \te^2 +3\te^4 \right )
\nonumber \\
&& -   \frac{1}{16}\left (\frac{GM}{c^2 \tp} \right )^3 \left \{ \left (197 - 205 \te^2 - 98 \te^4 -5 \te^6 \right )
- \chi^2 \left [ (4 + 6 \te^2)(1-3\cos^2 \ti ) - 9  \te^2 \sin^2 \ti \cos 2\tom \right ] \right \}
\nonumber \\
&&
- \frac{1}{4} \left (\frac{GM}{c^2 \tp} \right )^{7/2} \left [2(20+70 \te^2 +33\te^4) - \te^2 (9+4\te^2) \cos (2 \tom) \right ] \chi \cos \ti
\nonumber \\
&&
+ \frac{1}{128} \left (\frac{GM}{c^2 \tp} \right )^4 \left \{ \left (9923+12756 \te^2+11478 \te^4+952 \te^6+35 \te^8 \right )
+ \chi^2 \left [  8 (220 + 347 \te^2 +254 \te^4) 
\right .\right .
\nonumber \\
&&
 \left . \left .
\quad 
-4 (644 + 1099 \te^2 +695 \te^4)  \sin^2 \ti 
-8 \te^2 (94 + 80 \te^2 - 235 \sin^2 \ti - 177 \te^2 \sin^2 \ti )  \cos^2 \tom  \right ] \right \} \,,
\label{eq2:Econserved}
\\
L_e &=& \sqrt{m \tp}  \cos \ti \biggl \{ 1 + \frac{GM}{2c^2 \tp} (7 + \te^2) - \frac{1}{8} \left (\frac{GM}{c^2 \tp} \right )^2 
\left ( 37 -18 \te^2 -3\te^4 \right )
+ \frac{1}{16} \left (\frac{GM}{c^2 \tp} \right )^3  \biggl [411+481 \te^2 + 86 \te^4 + 5 \te^6 
\nonumber \\
&& \qquad \quad
+   \chi^2 \left [ \left ( 4(46 + 61 \te^2 ) - 5 (40 + 63 \te^2) \sin^2 \ti 
 -2 \te^2 (32 - 111 \sin^2 \ti ) \cos^2 \tom \right ) \right ] \biggr ]\biggr \}
\nonumber \\
&& - \sqrt{m \tp}  \chi  \biggl \{ \left (\frac{GM}{c^2 \tp} \right )^{3/2} (1+ \cos^2 \ti ) 
+ \frac{1}{4} \left (\frac{GM}{c^2 \tp} \right )^{5/2} \left [ 4(5+12 \te^2) - (30+41 \te^2) \sin^2 \ti
\right .
\nonumber \\
&&
\left.
\qquad \quad
- 2 \te^2 (4 - 5 \sin^2 \ti ) \cos^2 \tom \right ] \biggr \} \,,
\label{eq2:Leconserved}
 \\
C &=& m \tp \biggl \{ 1  + \frac{GM}{c^2 \tp} (7 + \te^2) - 4 \left (\frac{GM}{c^2 \tp} \right )^{3/2} \chi \cos \ti 
 + \frac{1}{2} \left (\frac{GM}{c^2 \tp} \right )^2 \left [ 6 + 16 \te^2 +2 \te^4 + \chi^2 \sin^2 \ti \left (1-2 \te^2 \cos^2 \tom \right ) \right ]
 \nonumber \\
&& - 2  \left (\frac{GM}{c^2 \tp} \right )^{5/2} \chi \cos \ti \left ( 12 + 13 \te^2 - 2\te^2 \cos^2 \tom \right )
+ \frac{1}{8} 
\left (\frac{GM}{c^2 \tp} \right )^3 \biggl [ 152 +570 \te^2 +125 \te^4 + 8 \te^6 
 \nonumber \\
&&
\qquad \quad
+ \chi^2 \left [ 4(54 + 61 \te^2) - (280 + 347 \te^2 )\sin^2 \ti - 2 \te^2 (32 - 49 \sin^2 \ti +4\te^2\sin^2 \ti )\cos^2 \tom \right ]\biggr ] \biggr \} \,.
\label{eq2:Cconserved}
\end{eqnarray} 
\label{eq2:ELeCconserved}
\end{subequations}
Thus $E$, $L_e$ and $C$ are constant over an orbital timescale.  However, a closer inspection of $E$ and $C$ indicates a potential problem.  Because $\tp$, $\te$ and $\ti$ are constant through 2.5PN order in the conservative part of the dynamics [see Eqs.\ (\ref{eq:dpdtheta}), (\ref{eq:didtheta}) and (\ref{eq:dedtheta2})], the PN corrections in $E$, $L_e$ and $C$ are constant through 3PN order, except for two: the 2PN terms in $E$ and $C$, proportional to $\chi^2$, that depend on $\tom$.  The quadrupole moment of the black hole is responsible for these $\tom$-dependent terms.  The pericenter $\tom$ varies at 1PN order at a rate $d\tom/d\theta = 3GM/c^2 \tp$, suggesting that $E$, and $C$ might not really be constant through 3PN order.  However, the variations induced by the advance of $\tom$ in those 2PN terms in $E$ and $C$ are {\em exactly cancelled} by the 3PN variations in $\tp$ and $\te$ [Eqs.\ (\ref{eq:dpdtheta}) and (\ref{eq:dedtheta2})] acting on the Newtonian terms $-GM(1-\te^2)/c^2 \tp$ and $m\tp$, respectively.  In $L_e$, the 3PN variations in $\tp$ and $\ti$ acting on the Newtonian term $(GM\tp)^{1/2} \cos \ti$ automatically cancel each other.  Thus $E$, $L_e$ and $C$ are also conserved on a pericenter precession timescale, as expected.  But to verify this, it was {\em essential} to include the relevant quadrupole-monopole cross-terms in the equations of motion, and to include the higher-order corrections in solving the Lagrange planetary equations.  This is an example of how the ``quadrupole conundrum'', discussed in \cite{2014PhRvD..89d4043W}, is resolved.

We also need the orbital period, given by
\begin{eqnarray}
P \equiv 2\pi  \left \langle \frac{dt}{d\phi} \right \rangle  
 =2\pi \left \langle \frac{(r^2/h)}{1-(r^2/h) \cos \iota Q^{(t)}_\Omega } \right \rangle \,.
 \label{eq2:perioddef}
\end{eqnarray}
Substituting the full solutions (\ref{eq2:ansatz}) for the orbital elements, we obtain, to 2PN order,
\begin{eqnarray}
P &=& 2\pi  \left ( \frac{\tp^3}{GM (1-\te^2)^3} \right )^{1/2}  \biggl [ 1 + \frac{3}{2} \frac{GM}{c^2 \tp} \left (\frac{4+ 9\te^2+2\te^4}{1-\te^2} \right )  + 6 \left ( \frac{GM}{c^2 \tp} \right )^{3/2} \chi \cos \ti
\nonumber \\
&& \quad - \frac{3}{16} \left ( \frac{GM}{c^2 \tp} \right )^{2} \frac{1}{(1-\te^2)^2} \left \{ 56-1214 \te^2-941 \te^4-151 \te^6 -40(1-\te^2)^{7/2} 
\right .
\nonumber \\
&& \left .\qquad \quad + \chi^2 (1-\te^2)(24-4 \te^2 - (32 + 7\te^2 - 18\te^2 \cos^2 \tom ) \sin^2 \ti \right \} \biggr ] \,.
\label{eq:period}
\end{eqnarray}
Higher-order contributions to $P$ will not be needed.
 
\subsection{Secular evolution with radiation reaction}
\label{sec:reaction}

We now include the radiation-reaction terms in the equations of motion.   At zeroth order in our two-scale analysis, we have the usual orbital average of the $Q^0_\alpha$, holding the orbit elements fixed on the right-hand-side.  But since we are working to 4.5PN order, we must again include higher-order corrections in the two-scale analysis.  Because we work only to linear order in the reduced-mass parameter $\eta$, we need to include only cross terms between the conservative and radiation-reaction sectors.  For example, periodic terms induced by 1PN, 1.5PNSO and 2PN conservative perturbations in the orbit elements occurring in the 2.5PN radiation reaction expressions could generate secular contributions at 3.5PN, 4PNSO and 4.5PN orders, while periodic terms induced by 2.5PN radiation-reaction perturbations in the orbit elements occurring in the 1PN, 1.5PNSO, and 2PN conservative expressions could also generate secular contributions at the same order.   These cross-term effects are why it is essential to include conservative terms at high PN orders when analysing radiation reaction effects at high PN orders.

We carry out the iterations of the two-scale equations to an order consistent with the 4.5PN order of the radiation reaction terms in the equations of motion.  After transforming from $\tal$ and $\tbe$ to $\te$ and $\tom$, we obtain
\begin{subequations}
\begin{eqnarray}
\frac{d\tp}{d\theta}_{\rm RR} &=& - \frac{8}{5} \eta \tp \left (\frac{GM}{c^2 \tp} \right )^{5/2} \left (8+7\te^2 \right ) + \frac{1}{210} \eta \tp \left (\frac{GM}{c^2 \tp} \right )^{7/2} \left ( 22072 - 6064 \te^2 - 1483 \te^4 \right )
\nonumber
\\
&& + \frac{2}{15} \eta \tp \left (\frac{GM}{c^2 \tp} \right )^4 \chi \cos \ti \left (968 + 2280 \te^2 + 297 \te^4 \right ) 
\nonumber 
\\
&&
- \frac{1}{11340} \eta \tp \left (\frac{GM}{c^2 \tp} \right )^{9/2} \left ( 8272600 + 777972 \te^2 - 947991 \te^4 - 4743 \te^6 \right ) \,,
\label{eq:dpdthetaRR} 
 \\
\frac{d\te}{d\theta}_{\rm RR} &=& - \frac{1}{15} \eta \te \left (\frac{GM}{c^2 \tp} \right )^{5/2} \left (304+121\te^2 \right ) + \frac{1}{840} \eta \te \left (\frac{GM}{c^2 \tp} \right )^{7/2} \left ( 144392 - 34768 \te^2 - 2251 \te^4 \right )
\nonumber 
\\
&& + \frac{1}{30} \eta \te \left (\frac{GM}{c^2 \tp} \right )^4 \chi \cos \ti \left (9400 + 10548 \te^2 + 789 \te^4 \right ) 
\nonumber 
\\
&&
- \frac{1}{30240} \eta \te \left (\frac{GM}{c^2 \tp} \right )^{9/2} \left ( 43837360 + 4062372 \te^2 - 1866490 \te^4 - 250065 \te^6 \right ) \,,
\label{eq:dedthetaRR}  
\\
\frac{d\ti}{d\theta}_{\rm RR} &=& -\frac{1}{30} \eta \left (\frac{GM}{c^2 \tp} \right )^4 \chi \sin \ti  \left [ 88 -1248 \te^2- 441 \te^2 + 24 \te^2 (20 + 13 \te^2) \cos^2 \tom \right ] \,,
\label{eq:didthetaRR}  
\\
\frac{d\tom}{d\theta}_{\rm RR} &=& \frac{4}{5} \eta \left (\frac{GM}{c^2 \tp} \right )^4 \chi \cos \ti \,\te^2 (20 + 13 \te^2) \sin \tom \cos \tom \,,
\label{eq:domdthetaRR}  
\\
\frac{d\tOm}{d\theta}_{\rm RR} &=& -\frac{4}{5} \eta \left (\frac{GM}{c^2 \tp} \right )^4 \chi  \te^2 (20 + 13 \te^2)  \sin \tom \cos \tom \,.
\label{eq:dOmdthetaRR}
\end{eqnarray}
\label{eq:dXdthetaRR} 
\end{subequations}
Notice that the 4PN spin-orbit terms act to reduce the orbital decay for prograde orbits ($\ti < \pi/2$) and act to enhance the decay for retrograde orbits ($\ti > \pi/2$).  This phenomenon has also been seen in numerical simulations of the late stage of inspiral of spinning black holes.  

Substituting Eqs.\ (\ref{eq:dXdthetaRR}) into Eqs.\ (\ref{eq2:ELeCconserved}), keeping contributions through 2PN order beyond the leading terms, and dropping $\chi^2$ terms, since they are not present in the radiation reaction terms, we obtain the rates of change of $E$, $L_e$ and $C$:
\begin{subequations}
\begin{eqnarray}
\frac{dE}{d\theta} &=& - \frac{32}{5} \eta \left ( \frac{GM}{c^2 \tp} \right )^{7/2} \biggl \{ \left ( 1 + \frac{73}{24} \te^2 + \frac{37}{96} \te^4 \right )
\nonumber \\
&& \quad 
- \left ( \frac{GM}{c^2 \tp} \right ) \left (\frac{95216+73224 \te^2-18074 \te^4-2393 \te^6}{5376} \right )
\nonumber \\
&& \quad
-  \left ( \frac{GM}{c^2 \tp} \right )^{3/2} \chi \cos \ti  \left (\frac{1936+ 12024 \te^2 +6582 \te^4 +195 \te^6}{192} \right )
\nonumber \\
&& \quad
+ \left ( \frac{GM}{c^2 \tp} \right )^2  \left (\frac{121274560+71538080 \te^2+191064 \te^4-317664  \te^6-427473 \te^8}{580608} \right )\biggr \} \,,
\label{eq:Edot}
\\
\frac{dLe}{d\theta} &=&  - \frac{32}{5} \eta \sqrt{GM\tp} \left ( \frac{GM}{c^2 \tp} \right )^{5/2} \biggl \{ \left ( 1 + \frac{7}{8} \te^2  \right ) \cos \ti
\nonumber \\
&& \quad 
- 5  \left ( \frac{GM}{c^2 \tp} \right ) \left ( \frac{6296-1000 \te^2-739 \te^4}{2688} \right ) \cos \ti
\nonumber \\
&& \quad 
- \frac{1}{192}\left ( \frac{GM}{c^2 \tp} \right )^{3/2} \chi \left [ 2(584+ 1944 \te^2 + 297 \te^4) - 3(488 + 1744 \te^2 +293 \te^4 ) \sin^2 \ti 
\right .
\nonumber \\
&&
\left .
 \quad \qquad 
+ 12 \te^2 (20 + 13 \te^2) \cos 2\tom \sin^2 \ti \right ]
\nonumber \\
&& \quad 
+ \left ( \frac{GM}{c^2 \tp} \right )^{2} \left ( \frac{14458192-819888 \te^2+38886 \te^4+147537 \te^6}{145152} \right )  \cos \ti 
 \biggr \} \,,
\label{eq:Ledot}
\\
\frac{dC}{d\theta} &=& - \frac{64}{5} \eta (GM\tp) \left ( \frac{GM}{c^2 \tp} \right )^{5/2} \biggl \{ \left ( 1 + \frac{7}{8} \te^2  \right ) 
\nonumber \\
&& \quad 
-  \left ( \frac{GM}{c^2 \tp} \right ) \left ( \frac{22072-14576 \te^2-4871 \te^4}{2688} \right ) 
\nonumber \\
&& \quad 
-  \left ( \frac{GM}{c^2 \tp} \right )^{3/2} \chi \cos \ti \left ( \frac{776+2112 \te^2+297 \te^4}{96} \right )
\nonumber \\
&& \quad 
+ \left ( \frac{GM}{c^2 \tp} \right )^{2} \left ( \frac{7837144-985668 \te^2+1212441 \te^4+294930 \te^6}{145152} \right ) 
 \biggr \} \,.
 \label{eq:Cdot}
\end{eqnarray}
\label{eq:ELeCdot}
\end{subequations}

\end{widetext}

\subsection{Comparison with other work}
\label{sec:compare}

We compare these results with other approaches in situations where there is overlap.  
Mora and Will \cite{2004PhRvD..69j4021M},  found the secular evolutions of the orbit elements in the conservative sector to 3PN order in the spinless ($\chi = 0$), arbitrary mass case, using the $\alpha-\beta$ formulation and the same two-scale approach.  In that limit, our Eqs.\ (\ref{eq:dXdtheta}) and (\ref{eq:dXdtheta2}), reduce to
$d\tp/d\theta = d\te/d\theta = d\ti/d\theta = d\tOm/d\theta = 0$, and 
\begin{equation}
\frac{d\tom}{d\theta} = 3u - \frac{3}{4} u^2 (10 - \te^2) + \frac{3}{2} u^3 (29 + 34 \te^2 ) \,,
\end{equation}
where $u = GM/c^2 \tp$.  On the other hand, Mora and Will obtained [Eq.\ (2.29) of \cite{2004PhRvD..69j4021M}]
\begin{equation}
\frac{d\tom}{d\theta} = 3\tilde{u} - \frac{3}{4} \tilde{u}^2 (10 - \te^2) + \frac{3}{2} \tilde{u}^3 (29 - 30 \te^2 ) \,.
\end{equation}
where $\tilde{u}$ was defined to be $\langle GM/c^2 p \rangle$.  These expressions appear to differ. However, $u$ and $\tilde{u}$ are not quite the same.   Using our solutions for $Y_p$ to the appropriate order, it is straighforward to show that $\tilde{u} = \langle GM/c^2 (\tp +  Y_p) \rangle = u + 32 \te^2 u^3 + O(u^5)$, and thus that the two expressions are in fact equivalent.  This illustrates the caution that, when working to high orders in perturbation theory, the precise definition of ``average'' is important.
Mora and Will also obtained the radiation-reaction driven evolution of the elements to 3.5PN order in the 
 spinless black hole case.   Eqs.\ (\ref{eq:dXdthetaRR}) agree completely with  Eq.\ (2.28) of \cite{2004PhRvD..69j4021M} to the relevant order; in this case the transformation between $\tilde{u}$ of Mora-Will and our $u =GM/c^2 \tp$ is a higher-order effect.  

Our results also agree with calculations of the rate of evolution of the orbital frequency $\Omega_{\rm circ}$ of a quasicircular inspiral orbit using high PN-order calculations of the energy flux combined with the assumption of energy balance \cite{1995PhRvD..52..821K,1995PhRvL..74.3515B}.  Through 2PN order beyond the quadrupole approximation (corresponding to our 4.5PN order in the equations of motion), including the leading spin-orbit contributions, the result in the small-$\eta$ limit is [Eq.\ (3) of \cite{1995PhRvL..74.3515B}] 
\begin{eqnarray}
\frac{d\Omega_{\rm circ}}{dt} &=& \frac{96}{5} \eta \left ( \frac{c^3}{GM} \right )^2 x^{11/3} \left [ 1 - \frac{743}{336} x^{2/3} 
\right .
\nonumber \\
&& \quad
\left .
- \frac{113}{12} x \chi \cos \ti + \frac{34103}{18144} x^{4/3} \right ] \,,
\label{eq:domegadt}
\end{eqnarray}
where $x \equiv GM \Omega_{\rm circ}/c^3$.
In \cite{1995PhRvL..74.3515B}, $\Omega_{\rm circ}$ was defined to be
\begin{eqnarray}
\Omega_{\rm circ} &\equiv& {|v|}/{r} 
\nonumber \\
&=& \frac{(GM)^{1/2}}{p^{3/2}} (1+\alpha \cos \phi + \beta \sin \phi)  (1+ \alpha^2 
\nonumber \\
&& \quad
+ \beta^2 + 2 \alpha \cos \phi + 2 \beta \sin \phi )^{1/2} \,.
\end{eqnarray}
Substituting our solutions for the orbit elements $p$, $\alpha$ and $\beta$ including periodic terms, and taking the $\te = 0$ limit, we obtain
\begin{eqnarray}
x &=&  \left ( \frac{GM}{c^2 \tp} \right )^{3/2} \left [ 1 - 6 \frac{GM}{c^2 \tp} - 4 \left (\frac{GM}{c^2 \tp} \right )^{3/2} \chi \cos \ti 
\right .
\nonumber \\
&& \qquad
\left .
+ 39 \left (\frac{GM}{c^2 \tp} \right )^2 \right ] \,.
\label{eq:xdef}
\end{eqnarray}
For $\te = 0$, the orbital period (\ref{eq:period}) becomes
\begin{eqnarray}
P &=& 2\pi \left ( \frac{\tp^3}{GM} \right )^{1/2} \left [ 1 + 6 \frac{GM}{c^2 \tp} + 6 \left (\frac{GM}{c^2 \tp} \right )^{3/2} \chi \cos \ti 
\right .
\nonumber \\
&& \qquad
\left .
- 3 \left (\frac{GM}{c^2 \tp} \right )^2 \right ] \,,
\end{eqnarray}
after dropping the 2PN contribution involving $\chi^2$.
Then calculating $d\Omega_{\rm circ}/dt = (2\pi/P)(c^3/GM) dx/d\theta$, inserting $(d\tp/d\theta)_{\rm RR}$ from Eq.\ (\ref{eq:dXdthetaRR}) in the $\te = 0$ limit, and then using Eq.\ (\ref{eq:xdef}) to convert the result from $\tp$ back to $x$, we obtain exactly Eq.\ (\ref{eq:domegadt}). 

Gergely {\em et. al.}  \cite{1998PhRvD..57..876G} derived the 4PN spin-orbit effects on the radiation-reaction evolution of the orbit elements by calculating fluxes of energy and angular momentum at infinity including the contributions from the current multipole moments of a rotating body (not necessarily a black hole), and including 1.5PN spin-orbit effects in the conservative equations of motion, while Sago and Fujita \cite{2015PTEP.2015g3E03S} calculated the fluxes in black-hole perturbation theory and then expanded in a PN sequence.   Both papers used a specific definition of the orbit elements $a_0$, $e_0$ and $i_0$, given by [see for example Eqs.\ (2.15) and (5.1) of \cite{1998PhRvD..57..876G}] 
\begin{eqnarray}
r_{\rm max(min)} &\equiv& a_0 (1 \pm e_0 ) \,,
\nonumber \\
cos(i_0) &=& \frac{L_e}{L} \,,
\end{eqnarray}
and calculated $\langle dE/dt \rangle$, $\langle dL_e/dt \rangle$, $\langle dL/dt \rangle$ in terms of those orbit elements.
Using our solutions (\ref{eq2:zerothorderavfree}) through 1.5PN order  for the orbital elements (including the periodic contributions), and using the approximation $L = \sqrt{C}$, it is straightforward to show that    
\begin{eqnarray}
a_0 &=& \frac{\tp}{1-\te^2} +  \left ( \frac{GM}{c^2} \right ) \frac{3+13 \te^2 - \te^4}{(1-\te^2)^2} 
\nonumber \\
&& \quad
+2 \tp \left ( \frac{GM}{c^2 \tp} \right )^{3/2} \chi \cos \ti + \dots \,,
\nonumber \\
e_0 &=& \te + \te  \frac{GM}{2c^2 \tp} (17 - 2\te^2) 
\nonumber \\
&& \quad
 - 2 \te (1-\te^2)  \left ( \frac{GM}{c^2 \tp} \right )^{3/2} \chi \cos \ti + \dots \,,
\nonumber \\
\cos i_0 &=& \cos \ti - \left ( \frac{GM}{c^2 \tp} \right )^{3/2} \chi \sin^2 \ti  + \dots\,.
\end{eqnarray}
Converting from $d/dt$ to $d/d\theta$ using the corrections to the orbital period in Eq.\ (\ref{eq:period}) through 1.5PN order, it is straightforward to show that the Newtonian and the 1.5PN spin-orbit contributions to the fluxes of $E$, $L_e$ and $L$ displayed in Eqs.\ (5.6), (5.7) and (5.8) of \cite{1998PhRvD..57..876G} and in Eqs.\ (32) -- (34) of \cite{2015PTEP.2015g3E03S} are completely equivalent to the corresponding contributions in our Eqs.\ (\ref{eq:ELeCdot}).  On the other hand, the 1PN and 2PN contributions in Eqs.\ (32) -- (34) of \cite{2015PTEP.2015g3E03S} are not equivalent.  This is undoubtedly another example of how different kinds of averages can affect higher-order PN terms.
Furthermore, at leading PN order, $\langle dX/dt \rangle$ is the same as $\langle dX/d\phi \rangle /\langle dt/d\phi \rangle$.   This also holds for the leading contribution of a specific type of effect, such as the spin-orbit contribution.  But because of ``cross-term'' effects, it does not necessarily hold for 1PN and 2PN  corrections to the leading Newtonian flux.    

We have not attempted a direct comparison with other related results \cite{1997CQGra..14.2357R,2008PhRvD..77f4035A,2009PhRvD..80l4018A,2016PhRvD..93f4058F,2016arXiv160908268G}, largely because they use rather different orbital parametrizations than those used here.

\section{Highly eccentric orbits}
\label{sec:eccentricorbits}

In this section, we prepare the groundwork for the numerical analysis of the evolution of highly eccentric orbits around a massive black hole,  to be addressed in Sec.\ \ref{sec:application}.   We have already noted the unexpected behavior of $r^{-1}$ in the $\te \to 1$ limit in Eq.\ (\ref{eq:rinve1}).
Examination of the expression for $E$ [Eq.\ (\ref{eq2:Econserved})] reveals that, for fixed semilatus rectum $\tp$,  the Newtonian energy vanishes as the eccentricity $\te$ tends toward unity, but the PN corrections remain finite.  Similarly, while the Newtonian part of the period $P$ [Eq.\ (\ref{eq:period})]  tends to infinity in this limit as $(1-\te^2)^{-3/2}$ as expected for the transition from a bound to an unbound orbit, the PN and 2PN  corrections to the period blow up even faster.   It suggests that the true boundary between bound and unbound orbits in PN theory corresponds to something other than $\te = 1$.   This behavior is reminiscent of the singular behavior of the $\te-\tom$ parametrization of osculating orbits in the limit of circular orbits in PN theory, which forces either the $\alpha-\beta$ parametrization of the orbit elements or a parametrization in terms of two eccentricities \cite{1985AIHS...43..107D,1986AIHS...44..263D}.  But to our knowledge, this behavior at $\te = 1$ does not seem to have received much attention.   

In an attempt to cure this problem,  we shall define a post-Newtonian corrected eccentricity $e$, subject to the following constraints:  (i)  the limit $e \to 0$ still corresponds to $\te \to 0$; (ii) the energy vanishes as $(1-e^2)$ as $e \to 1$, holding $\tp$ fixed.    We define the transformation to the new eccentricity by

\begin{widetext}

\begin{eqnarray}
\te &\equiv&e \left [ 1 + \frac{GM}{c^2 \tp} f_1(e) + \left (\frac{GM}{c^2 \tp} \right )^{3/2} f_{1.5}(e) \chi \cos \ti  +  \left (\frac{GM}{c^2 \tp} \right )^2  f_2(e)
\right .
\nonumber \\
&& \qquad
\left .
 + \left (\frac{GM}{c^2 \tp} \right )^{5/2} f_{2.5}(e) \chi \cos \ti 
+  \left (\frac{GM}{c^2 \tp} \right )^3  f_3(e) + \dots \right ] \,,
\label{eq:etransform}
\end{eqnarray}
where $f_\alpha (e)$ are to be polynomials in $e$, constrained to be finite as $e \to 0$.  We substitute this  into Eq.\ (\ref{eq2:Econserved}) and require that  $E$ acquire the form

\begin{eqnarray}
E &\to& -\frac{1}{2} \frac{GM (1-e^2)}{\tp} \left [ 1 + \frac{GM}{c^2 \tp} g_1(e) + \left (\frac{GM}{c^2 \tp} \right )^{3/2} g_{1.5}(e) \chi \cos \ti +  \left (\frac{GM}{c^2 \tp} \right )^2  g_2(e) 
\right .
\nonumber \\
&& \qquad
\left .
+ \left (\frac{GM}{c^2 \tp} \right )^{5/2} g_{2.5}(e) \chi \cos \ti +  \left (\frac{GM}{c^2 \tp} \right )^3  g_3
(e) + \dots \right ] \,.
\end{eqnarray}
The solutions are
\begin{eqnarray}
f_1(e) &=&  \frac{ 8 g_1(e) (1-e^2) - 19 - 38e^2 - 3e^4}{8e^2} \,,
\nonumber \\
f_{1.5}(e) &=& \frac{g_{1.5}(e) (1-e^2)}{e^2} \,,
\nonumber \\
f_{2}(e) &=& \frac{1}{e^4} \biggl \{ (8 g_1(e)-19)^2 
- 4 e^2 \left [\left (8 g_1(e) -19\right ) \left (4 g_1(e) -19\right )+32 g_2(e) + 394 - 8 \chi^2 \left (1-3 \cos^2 \ti \right ) 
\right ] +O(e^4)  \biggr \} \,,
    \nonumber \\
f_{2.5}(e) &=& -\frac{1}{8e^4} \biggl \{ (8g_1(e) -19 )g_{1.5}(e) - e^2 \left [  g_{1.5}(e) (16 g_1(e)-57) +80 + 8g_{2.5}(e) \right ]  + O(e^4)  \biggr \}\,,
\nonumber \\
f_3(e) &=& \frac{1}{1024 \, e^6} \biggl \{ (8 g_1(e)-19)^3  -2 e^2 \left [  (8 g_1(e) -19)^2 (12 g_1(e) -19 )  
\right .
\nonumber \\
&& \qquad
\left .
+ 4(8 g_1(e) -19) \left (16 g_2(e) + 197 
-4 \chi^2 (1-3 \cos^2 \ti ) \right )  +256 g_{1.5}(e)^2 \chi^2 \cos^2 \ti \right ] 
\nonumber \\
&& \qquad
+ e^4 \left [ 4 (8 g_1(e) -19)^2 (6 g_1(e) -7) +8 (16 g_1(e) -57)\left (16 g_2(e) + 197 -4 \chi^2  (1-3 \cos^2 \ti ) \right ) 
\right .
\nonumber \\
&& \qquad
\left .
+ (8 g_1(e) -19) \left ( 755 + \chi^2 (32 - 120 \sin^2 \ti + 144 \cos^2 \tom \sin^2 \ti ) \right )
\right .
\nonumber \\
&& \qquad
\left .
+1024 g_3(e) -79384 + 6528 \chi^2 - 20608 \chi^2 \cos^2 \ti + 1024 g_{1.5}(e)^2 \chi^2 \cos^2 \ti 
\right ] + O(e^6) 
\biggr \} \,,
 \end{eqnarray}
A necessary and sufficient condition  that the $f_\alpha (e)$ be regular at $e=0$ is the set of constraints
\begin{eqnarray}
g_1(e) &=& \frac{19}{8}  \,,
\nonumber \\
g_{1.5}(e) &=&  0 \,,
\nonumber \\
g_2(e) &=& -\frac{197}{16} + \frac{1}{4} \chi^2 (1 - 3 \cos^2 \ti )  \,,
\nonumber \\
g_{2.5}(e) &=&  -10 \,,
\nonumber \\
g_3(e) &=& \frac{9923}{128} - \frac{1}{8} \chi^2 (51 - 161 \cos^2 \ti )  \,,
\end{eqnarray}
resulting in the transformation
\begin{eqnarray}
\te &=& e \biggl \{ 1 - \frac{3}{8} \frac{GM}{c^2 \tp} \left ( 19 +e^2 \right )
    +  \frac{1}{128} \left (\frac{GM}{c^2 \tp} \right )^2  \left [ 5351 + 698 e^2+23 e^4 + 8 \chi^2 \left (20 -  \sin^2 \ti (39-18 \cos^2 \tom ) \right )\right ] 
    \nonumber \\
    && \qquad
 +  \frac{1}{4} \left (\frac{GM}{c^2 \tp} \right )^{5/2} \chi \cos \ti \left [7(10+27 e^2) - 2(9+4e^2) \cos^2 \tom \right ]
  \nonumber \\
    && \qquad
  -  \frac{1}{1024} \left (\frac{GM}{c^2 \tp} \right )^{3} \left [ 343065  +107609 e^2 + 4243 e^4 + 91 e^6
  \right .
  \nonumber \\
  && \qquad \quad
  \left .
  + \chi^2 \left (32( 1571+571 e^2) - 8(10791+3293 e^2) \sin^2 \ti - 128 (47+ 40 e^2) \cos^2 \tom
   \right . 
   \right .
  \nonumber \\
  && \qquad \qquad
  \left .
  \left .
  +16 (2137+843 e^2) \cos^2 \tom  \sin^2 \ti \right ) \right ]
     \biggr \}\,,
       \label{eq4:etransform}
\end{eqnarray}
With this transformation, $E$ now takes the form
\begin{eqnarray}
c^{-2} E &=& -\frac{1}{2} \frac{GM (1-e^2)}{c^2 \tp} \left [ 1 - \frac{19}{4} \frac{GM}{c^2 \tp}  +  \frac{1}{8}\left (\frac{GM}{ c^2 \tp} \right )^2 \left (197 - 4 \chi^2 (1 - 3 \cos^2 \ti ) \right ) 
 \right .
  \nonumber \\
  && \qquad 
  \left .
  + 20 \left (\frac{GM}{ c^2 \tp} \right )^{5/2} \chi \cos \ti
  - \frac{1}{64} \left (\frac{GM}{ c^2 \tp} \right )^{3} \left ( 9923 + 16 \chi^2 (110 - 161 \sin^2 \ti ) \right )
  \right ] \,.
  \label{eq4:Enew}
\end{eqnarray}
As expected, in the limit $e \to 0$, $E$ agrees with Eq.\ (\ref{eq2:Econserved}) in the limit $\te \to 0$.   

In addition, there is no longer any $\tom$ dependence in the 2PN terms in $E$, thus curing the ``quadrupole conundrum" discussed in Sec.\ \ref{sec:conservative}.   This can be understood as follows:  in the conservative dynamics,  $\tp$ and $\te$ are constant through 2.5PN order, but vary at 3PN order [see Eqs.\ (\ref{eq:dpdtheta}) and (\ref{eq:dedtheta2})].  However, $\tom$ appears in the 2PN term in the transformation between $\te$ and $e$ in Eq.\ (\ref{eq4:etransform}).  Thus through 3PN order
\begin{equation}
\frac{d\te}{d\theta} = \frac{de}{d\theta} - \frac{9}{4} e \left (\frac{GM}{ c^2 \tp} \right )^{2} \chi^2 \sin^2 \ti \sin \tom \cos \tom \frac{d\tom}{d\theta} \,.
\end{equation}
Substituting Eq.\ (\ref{eq:dedtheta2}) for $d\te/d\theta$ along with $d\tom/d\theta = 3GM/c^2 \tp$, we obtain
\begin{equation}
\frac{de}{d\theta} = 3 e \left (\frac{GM}{ c^2 \tp} \right )^{3} \chi^2 \sin^2 \ti \sin \tom \cos \tom \,.
\label{eq:dedthetanew}
\end{equation}
Combining this with Eq.\ (\ref{eq:dpdtheta}) for $d\tp/d\theta$, which is unaffected to this order by the change in $e$, we find that 
\begin{equation}
\frac{d \tilde{a}}{d\theta} \equiv \frac{d}{d\theta} \left (\frac{\tp}{(1-e^2)} \right ) = O \left  (\frac{GM}{ c^2 \tp} \right )^{7/2}  \,.
\end{equation}
In other words, 3PN variations in $\tilde{a}$ are no longer needed to compensate for variations in $E$ induced by the advance of the pericenter, since that orbit element no longer appears in $E$.

 The transformation of $e$ simultaneously cures the bad behavior of $P$, which now takes the form
\begin{eqnarray}
P &=& 2\pi \left ( \frac{\tp^3}{GM (1-e^2)^3} \right )^{1/2} \biggl \{ 1 + \frac{3}{8} \frac{GM}{c^2 \tp} (16 - 5e^2) +  6 \left (\frac{GM}{c^2 \tp} \right )^{3/2} \chi \cos \ti
\nonumber \\
&& \quad
 - \frac{3}{128} \left (\frac{GM}{c^2 \tp} \right )^2 \left [ 448 - 88 e^2 + 35e^4 - 320 (1-e^2)^{3/2} - 64 \chi^2 (1- 4\cos^2 \ti)  \right ] \biggr \} \,.
 \label{eq3:Pnew}
\end{eqnarray}
It also cures the behavior of the orbit shown in Eq.\  (\ref{eq:rinve1}); in terms of $e$, $r^{-1}$ now factorizes into
\begin{equation}
r^{-1} = \frac{1+\cos f}{\tp} \left [ 1 - 2 \frac{GM}{c^2 \tp} - \left (\frac{GM}{c^2 \tp} \right )^{3/2} \chi \cos \ti \left ( 6 - 2 \cos f - \sin f \sin 2\phi \right ) + \dots \right ]  \,.
\label{eq:rinve2}
\end{equation}

In terms of the new eccentricity, $C$ and $L_e$ become
\begin{subequations}
\begin{eqnarray}
L_e &=& \sqrt{m \tp} \biggl \{  \cos \ti  \biggl [1 + \frac{GM}{2c^2 \tp} (7 + e^2) - \frac{1}{8} \left (\frac{GM}{c^2 \tp} \right )^2 \left ( 37 +39e^2 \right )  
+ \frac{1}{16} \left (\frac{GM}{c^2 \tp} \right )^3 \left ( 411 +1043e^2 + 18 e^4 
\right .
\nonumber \\
&& \qquad
\left .
+ 2\chi^2 [ 92 + 132e^2 - (100+177 e^2) \sin^2 \ti 
-32 e^2 \cos^2 \tom + 120 e^2 \cos^2 \tom \sin^2 \ti ] \right ) 
\biggr ] 
\nonumber \\
&&
- \chi  \left (\frac{GM}{c^2 \tp} \right )^{3/2} \biggl \{ (1+ \cos^2 \ti ) 
+ \frac{GM}{4c^2 \tp}  \left [ 20 + 48e^2 - (30 + 41e^2) \sin^2 \ti 
\right .
\nonumber \\
&& \qquad
\left . 
-8e^2 \cos^2 \tom +10e^2\cos^2 \tom \sin^2 \ti \right ]
 \biggr \} \,,
 \label{eq3:Lenew}
\\
C &=& m \tp \biggl \{ 1  + \frac{GM}{c^2 \tp} (7 + e^2) - 4 \left (\frac{GM}{c^2 \tp} \right )^{3/2} \chi \cos \ti 
+ \frac{1}{4} \left (\frac{GM}{c^2 \tp} \right )^2 \left [ 12 - 25 e^2 + e^4 + 2\chi^2 \sin^2 \ti \left (1-2 e^2 \cos^2 \tom \right ) \right ]
 \nonumber \\
 &&  \quad
- 2 \left (\frac{GM}{c^2 \tp} \right )^{5/2} \chi \cos \ti  \left ( 12 + 13 e^2 - 2e^2 \cos^2 \tom \right )
+ \frac{1}{8} \left (\frac{GM}{c^2 \tp} \right )^{3} \left [ 152  + 733 e^2 -21 e^4 
\right.
 \nonumber \\
 && 
 \left .
 \qquad
 +2 \chi^2 \left ( 108 + 132e^2 - (140+193e^2)\sin^2 \ti -32 e^2 \cos^2 \tom + (115-e^2)e^2 \cos^2 \tom \sin^2 \ti \right ) \right ]
  \biggr \} \,.
   \label{eq3:Cnew}
\end{eqnarray}
 \label{eq3:ECnew}
\end{subequations}

The final step in transforming to the new eccentricity is to determine the evolution induced by radiation reaction.   We invert Eq.\ (\ref{eq4:etransform}) to obtain $e$ as a function of $\te$, $\tp$, $\ti$ and $\tom$, find $de/d\theta$ by inserting the radiation-reaction equations (\ref{eq:dXdthetaRR}) for the $d\tilde{X}_\alpha/d\theta$, and then convert back to the new $e$.   In the remaining $d\tilde{X}_\alpha/d\theta_{\rm RR}$ expressions, we convert from $\te$ to $e$.  The results are
\begin{subequations}
\begin{eqnarray}
\frac{d\tp}{d\theta}_{\rm RR} &=& - \frac{8}{5} \eta \tp \left (\frac{GM}{c^2 \tp} \right )^{5/2} \left (8+7e^2 \right ) 
+ \frac{1}{210} \eta \tp \left (\frac{GM}{c^2 \tp} \right )^{7/2} \left ( 22072 + 27452 e^2 +281 e^4 \right )
\nonumber 
\\
&& + \frac{2}{15} \eta \tp \left (\frac{GM}{c^2 \tp} \right )^4 \chi \cos \ti \left (968 + 2280 e^2 + 297 e^4 \right ) 
\nonumber 
\\
&&
- \frac{1}{810} \eta \tp \left (\frac{GM}{c^2 \tp} \right )^{9/2} \left ( 590900 + 941316 e^2 - 100860 e^4 - 4383 e^6 \right ) \,,
\label{eq:dpdthetaRRnew} 
\\
\frac{de}{d\theta}_{\rm RR} &=& - \frac{1}{15} \eta e \left (\frac{GM}{c^2 \tp} \right )^{5/2} \left (304+121e^2 \right ) 
+ \frac{1}{840} \eta e \left (\frac{GM}{c^2 \tp} \right )^{7/2} \left ( 221000 + 120086 e^2 + 1277 e^4 \right )
\nonumber 
\\
&& + \frac{1}{30} \eta e \left (\frac{GM}{c^2 \tp} \right )^4 \chi \cos \ti \left (9400 + 10548 e^2 + 789 e^4 \right ) 
\nonumber 
\\
&&
- \frac{1}{15120} \eta e \left (\frac{GM}{c^2 \tp} \right )^{9/2} \left ( 39598064 + 26131872 e^2 - 1139399  e^4 - 150795 e^6 \right ) \,,
\label{eq:dedthetaRRnew} 
\\
\frac{d\ti}{d\theta}_{\rm RR} &=& -\frac{1}{30} \eta \left (\frac{GM}{c^2 \tp} \right )^4 \chi \sin \ti  \left [ 88 -1008 e^2- 285 e^2 + 12 e^2 (20 + 13 e^2) \cos (2\tom) \right ] \,,
\label{eq:didthetaRRnew} 
\\
\frac{d\tom}{d\theta}_{\rm RR} &=& \frac{2}{5} \eta \left (\frac{GM}{c^2 \tp} \right )^4 \chi \cos \ti \,e^2 (20 + 13 e^2) \sin (2\tom) \,,
\label{eq:domdthetaRRnew}  
\\
\frac{d\tOm}{d\theta}_{\rm RR} &=& -\frac{2}{5} \eta \left (\frac{GM}{c^2 \tp} \right )^4 \chi  e^2 (20 + 13 e^2) \sin (2\tom) \,.
\label{eq:dOmdthetaRRnew}
\end{eqnarray}
\label{eq:dXdthetaRRnew}
 \end{subequations}

\end{widetext}

\section{Numerical evolutions}
\label{sec:application}

We now wish to apply the analytic results obtained in Secs.\ \ref{sec:eom} -- \ref{sec:eccentricorbits} to the long-term evolution of highly eccentric orbits around a massive spinning black hole.  For a $10^6 M_\odot$ black hole, these would be orbits with semimajor axis $a \sim$ milliparsecs, and $1-e \sim 10^{-5}$.  Since $GM/c^2$ for the black hole is of order 50 nanoparsecs, this corresponds to an initial semilatus rectum $p \sim 2a(1-e) \sim 20 (GM/c^2)$.  Since our PN expansion is in powers of $GM/c^2 \tp$, this is a regime where we might hope to obtain reasonable results for the orbital evolution.   

We first average the secular evolution equations (\ref{eq:dXdthetaRRnew}) over $\tom$; since the pericenter advances on a timescale that is short compared to the radiation-reaction timescale, this is a reasonable approximation (since the dependence on $\tom$ occurs only in the highest order terms in Eqs.\ (\ref{eq:dXdthetaRRnew}), we do not need to carry out an {\em additional} two-scale analysis, and can average over $\tom$ holding the other elements fixed. (Notice from Eqs.\ (\ref{eq:dXdtheta}) and (\ref{eq:dedthetanew}) that the 3PN conservative variations in $\tp$, $e$, and $\ti$ also average to zero over a pericenter timescale.)  We then scale $\tp$ by $GM/c^2$, defining the dimensionless variables $\epsilon \equiv GM/c^2 \tp_i$ and $x \equiv \tp/\tp_i$, where $\tp_i$ is the initial value of $\tp$.   In terms of these variables, the evolution equations become (henceforth, we drop the tildes on all our orbital variables)
\begin{subequations}
\begin{eqnarray}
\frac{dx}{d\theta}_{\rm RR} &=& - \frac{8}{5} \eta \epsilon^{5/2} x^{-3/2} \left (8+7e^2 \right ) 
\nonumber 
\\
&&
+ \frac{1}{210} \eta \epsilon^{7/2} x^{-5/2} \left ( 22072 + 27452 e^2 +281 e^4 \right )
\nonumber 
\\
&& + \frac{2}{15} \eta \epsilon^{4} x^{3} \chi \cos \iota \left (968 + 2280 e^2 + 297 e^4 \right ) 
\nonumber 
\\
&&
- \frac{1}{810} \eta \epsilon^{9/2} x^{-7/2} \left ( 590900 + 941316 e^2 
\right .
\nonumber 
\\
&&
\left .
\qquad 
- 100860 e^4 - 4383 e^6 \right ) \,,
\label{eq:dpdthetaRRscaled} 
\\
\frac{de}{d\theta}_{\rm RR} &=& - \frac{1}{15} \eta e \epsilon^{5/2} x^{-5/2} \left (304+121e^2 \right ) 
\nonumber 
\\
&&
+ \frac{1}{840} \eta e \epsilon^{7/2} x^{-7/2} \left ( 221000 + 120086 e^2 
\right .
\nonumber 
\\
&&
\left .
\qquad 
+ 1277 e^4 \right )
\nonumber 
\\
&& + \frac{1}{30} \eta e \epsilon^{4} x^{-4} \chi \cos \iota \left (9400 + 10548 e^2 + 789 e^4 \right ) 
\nonumber 
\\
&&
- \frac{1}{15120} \eta e \epsilon^{9/2} x^{-9/2} \left ( 39598064 + 26131872 e^2 
\right .
\nonumber 
\\
&&
\left .
\qquad 
- 1139399  e^4 - 150795 e^6 \right ) \,,
\label{eq:dedthetaRRscaled} 
\\
\frac{d\iota}{d\theta}_{\rm RR} &=& -\frac{1}{30} \eta \epsilon^{4} x^{-4}\chi \sin \iota  \left ( 88 -1008 e^2
\right .
\nonumber 
\\
&&
\left .
\qquad 
- 285 e^2  \right ) \,,
\label{eq:didthetaRRscaled} 
\\
\frac{d\tom}{d\theta}_{\rm RR} &=& 0 \,,
\label{eq:domdthetaRRscaled}  
\\
\frac{d\tOm}{d\theta}_{\rm RR} &=& 0 \,.
\label{eq:dOmdthetaRRscaled}
\end{eqnarray}
\label{eq:dXdthetaRRscaled}
 \end{subequations}
Notice that, when the initial $p$ is scaled in units of $GM/c^2$, and with $x_i = 1$ the orbital evolution as a function of phase $\theta$ depends only on $\epsilon$, $e_i$ and $\iota_i$ and is independent of the mass of the black hole.   We shall see that the only place where the black-hole mass plays a role is in setting the conversion from orbital phase to time.

In addition, the effect of spin-orbit radiation reaction on the inclination $\iota$ is so small that the inclination has been found to be constant in all our numerical solutions.  Henceforth we shall drop Eq.\ (\ref{eq:didthetaRRscaled}), and treat the inclination as strictly constant.  

\subsection{Capture by  the black hole}
\label{sec:capture}

As the body inspirals toward the black hole, its angular momentum decreases to a point whereupon, even in the absence of additional radiation reaction, it finds itself in an orbit with no inner turning point, and it ultimately crosses the event horizon. 
In most previous work this critical angular momentum has been chosen to be $4GM/c$, corresponding to the angular momentum of a particle with zero energy at the innermost unstable peak of the effective potential in the Schwarzschild geometry.   For rotating black holes, this is not a good approximation, because of the strong effects of frame dragging, among other things.  In Ref.\ \cite{2012CQGra..29u7001W}, we derived a general condition for capture by a rotating black hole, for a body with zero energy (or with relativistic energy $\tilde{E} =1$), by finding the critical value of the Carter constant such that the effective radial potential for a geodesic simultaneously vanish and have vanishing radial derivative (with a suitable sign for the second radial derivative), corresponding to a turning point right at the unstable peak of the potential.   Defining ${L} \equiv \sqrt{C}$, and $\cos \iota \equiv L_e/L$, the result was an eighth-order polynomial equation for the critical value $L_c$ as a function of $\iota$ and the Kerr parameter $\chi$.  The numerical results are plotted in Fig.\ 1 of \cite{2012CQGra..29u7001W}.   We then found an analytic approximation to the curves given by
\begin{equation}
{L}_c = \frac{2GM}{c} \left [ 1 +  \sqrt{1 - \chi \cos \iota - \frac{1}{8}\chi^2 \sin^2 \iota F(\chi, \cos \iota) } \right ] \,,
\label{Lcrit}
\end{equation}
which has the correct behavior both when $\chi = 0$ (Schwarzschild) and when $\iota = 0$ or $\iota = \pi$ (equatorial orbits) for arbitrary $\chi$.  The function $F$ is a power series in $\chi$
\begin{eqnarray}
F(\chi, \cos \iota) &=& 1 + \frac{1}{2} \chi \cos \iota + \frac{1}{64} \chi^2 (7 + 13 \cos^2 \iota) 
\nonumber \\
&&
+ \frac{1}{128}  \chi^3 \cos \iota (23 + 5 \cos^2 \iota)
\nonumber \\
&&
+ O( \chi^4) \,.
\label{Fseries}
\end{eqnarray}
This series solution agrees with the numerical solutions of the critical condition to better than $0.5$ percent for $0 \le \chi \le 0.9$; and to better than $5$ percent for $0.9 \le \chi \le 0.99$.   

In our numerical evolutions, we will impose the condition $\sqrt{C} = L_c$, where $C$ is given by Eq.\ (\ref{eq3:Cnew}) and $L_c$ is given by Eq.\ (\ref{Lcrit}).   However, in the capture condition (\ref{Lcrit}) the inclination was defined by $\cos \iota \equiv L_e/L$; from Eqs.\ (\ref{eq3:Lenew}) and  (\ref{eq3:Cnew}), we see that 
\begin{equation}
\cos \iota = \cos \ti - \left ( \frac{GM}{c^2 p} \right )^{3/2} \chi \sin^2 \ti + O\left (\frac{GM}{c^2 p} \right )^2  \,.
\end{equation}
The difference between $\iota$ and $\ti$ is a maximum for polar orbits, but then we expect ${GM}/{c^2 p}$ to be less than 1/16, so we expect the difference between the two measures of inclination to be only a few percent.  We will ignore that difference.

\begin{figure}[t]
\begin{center}

\includegraphics[width=3.5in]{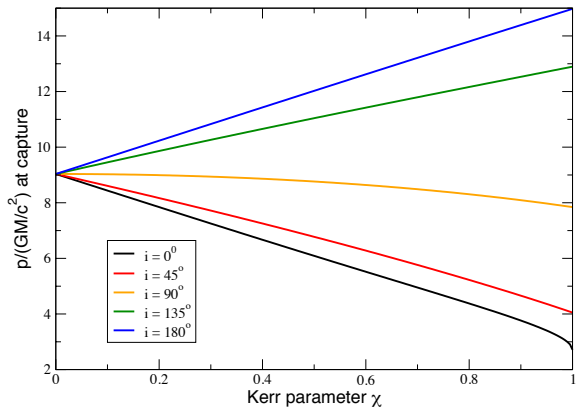}

\caption{\label{fig:pcapture}  Critical values of $p$ for capture by a rotating black hole as a function of the Kerr parameter $\chi$, for various orbital inclinations.  Orbital energy $E$ is assumed to be zero.}
\end{center}
\end{figure}

Using the expression for $C$ to 2PN order and converting to our variables $x$ and $\epsilon$, we obtain the capture condition
\begin{eqnarray}
&&x^{1/2} \left [ 1 + \frac{\epsilon}{2x} (7 +e^2) -2 \left ( \frac{\epsilon}{x} \right )^{3/2} \chi \cos \iota 
\right .
\nonumber \\
&& \qquad
\left .
- \frac{1}{8} \left ( \frac{\epsilon}{x} \right )^{2} \left (37 + 39e^2 - 2\chi^2 (1-e^2)\sin^2 \iota \right )\right ] 
\nonumber \\
&& \quad
 = \epsilon^{1/2} \left [ 1 + \sqrt{1 - \chi \cos \iota - \frac{1}{8}\chi^2 \sin^2 \iota F(\chi, \cos \iota) } \right ] \,.
 \nonumber \\
 \label{eq:condition}
\end{eqnarray}
Solutions for $p_c/(GM/c^2) = x_c /\epsilon$ as a function of $\iota$ and $\chi$ are shown in Fig.\ \ref{fig:pcapture}.   Since the orbits will have circularized to varying degrees by the time of plunge, we have chosen $e = 0$ in that figure; for $e=0.5$, for example, the curves are virtually the same.

We shall use the capture condition (\ref{eq:condition}) to terminate the numerical evolutions of $x$ and $e$.  An important caveat is that this condition strictly applies only for orbits with $E=0$; however, since $ E \sim GM/p$, we might expect the error in our capture condition to be $O(GM/c^2 p)$, smaller for retrograde orbits, larger for prograde orbits.    Improvement of the capture criterion for $E \ne 0$ is a topic for further work.
 
\subsection{Evolution in terms of orbital phase}
\label{sec:evolutionphase}

We have integrated Eqs.\ (\ref{eq:dpdthetaRRscaled}) and  (\ref{eq:dedthetaRRscaled}) numerically, for initial values of $p$ ranging from 10 to 100 $GM/c^2$, for $1-e_i$ ranging from $10^{-2}$ to $10^{-5}$, and for $\iota_i$ ranging from $0$ (equatorial prograde) to $\pi$ (equatorial retrograde).  We choose $\eta = 5 \times 10^{-5}$, corresponding to a $50 \, M_\odot$ object orbiting a $10^6 M_\odot$ black hole.   We will frequently choose $\chi =1$, in order to maximize the effects due to the spin of the black hole.

\begin{figure*}[t]
\begin{center}

\includegraphics[width=5in]{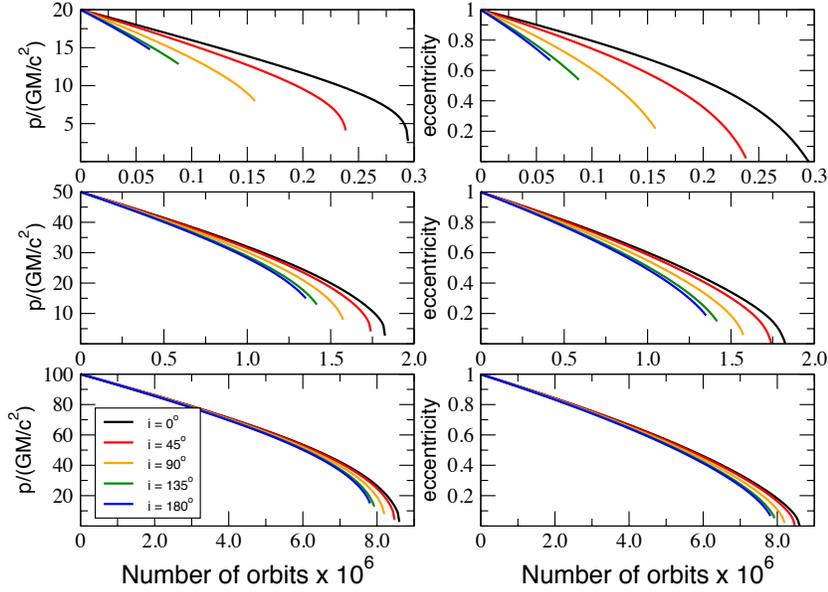}

\caption{\label{fig:pevsphase} Evolution of semilatus rectum $p$ and eccentricity $e$ vs. number of orbits, for three initial values of $p$; in all cases, $e_i = 0.999$ and $\chi =1$.   }
\end{center}
\end{figure*}

\begin{figure*}[t]
\begin{center}

\includegraphics[width=3.5in]{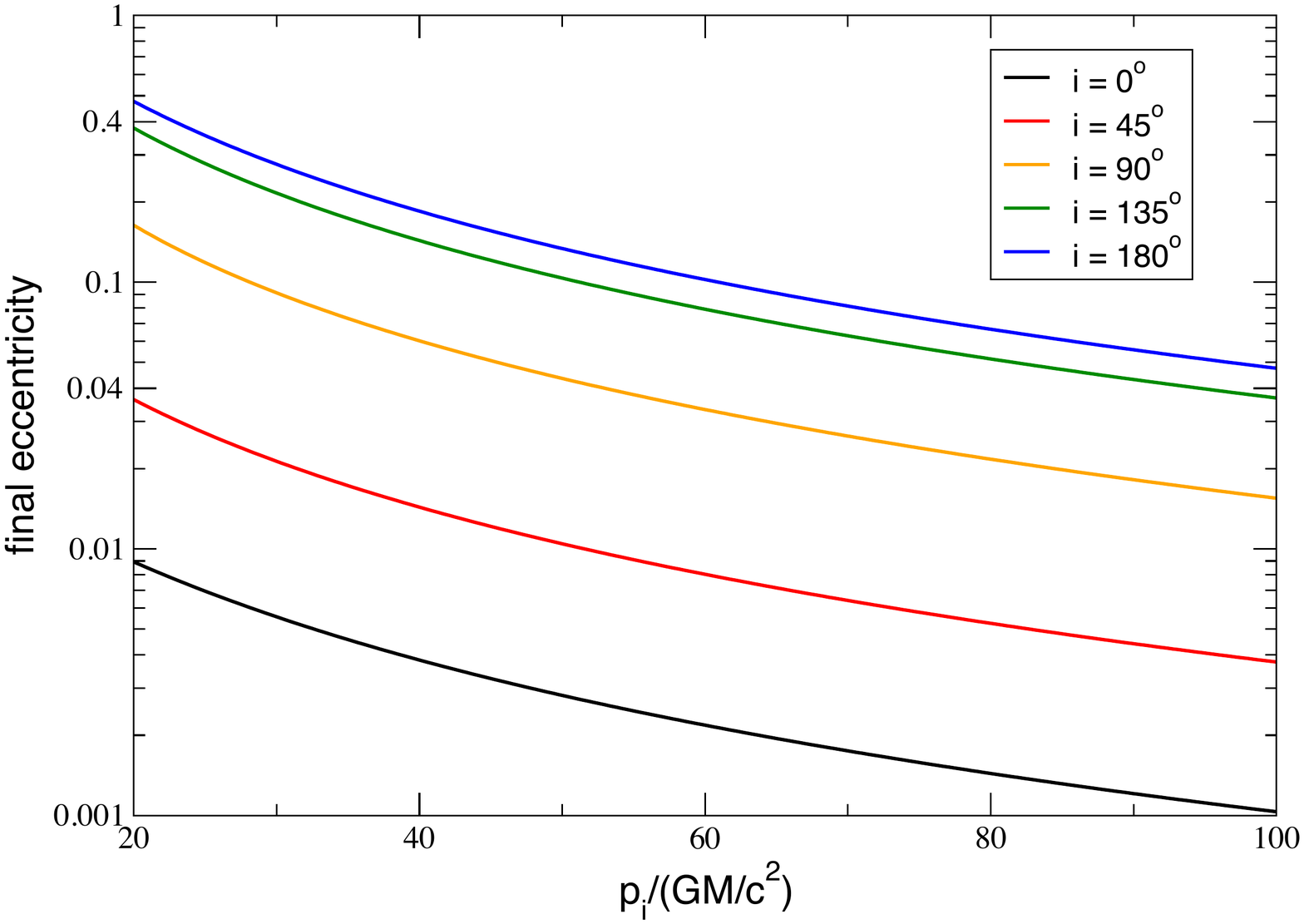}
\includegraphics[width=3.5in]{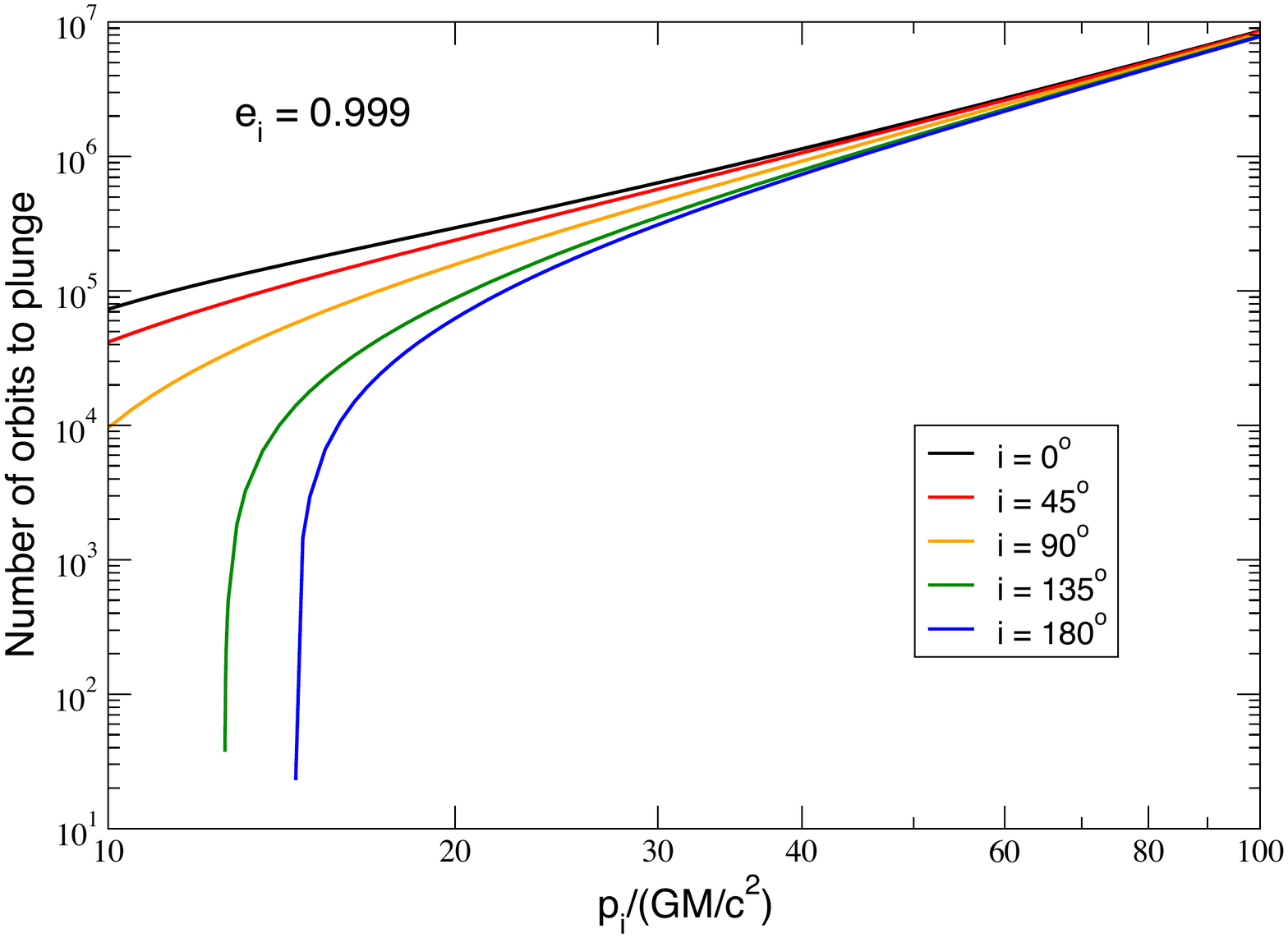}

\caption{\label{fig:efinal} Left panel: Orbital eccentricity at time of plunge vs. initial $p$.   Right panel: Number of orbits vs. initial $p$.  In both cases $e_i = 0.999$ and $\chi =1$.}
\end{center}
\end{figure*}

Figure \ref{fig:pevsphase} shows the evolution of $p/(GM/c^2)$ and $e$ as a function of the number of orbits, for initial values $p_i/(GM/c^2)$ 100, 50 and 20, and for $e_i = 0.999$.   We find, not surprisingly, that, for initial values of $e$ very near unity, the evolution with phase is independent of $e_i$.   Retrograde orbits are seen to decay faster than prograde orbits, a phenomenon that is clear from the spin-orbit terms in Eqs.\ (\ref{eq:dpdthetaRRscaled}) and (\ref{eq:dedthetaRRscaled}), and that has also been observed in numerical relativity simulations of inspiraling spinning black holes.  In some mysterious way, this may be related to the cosmic censorship hypothesis;  since any residual angular momentum that the particle carries into the black hole increases the hole's angular momentum if the inspiral orbit is prograde, the particle must inspiral for longer to ensure that enough angular momentum is radiated away to preclude creating a black hole with $\chi >1$ (a naked singularity).  No such issue is present for retrograde orbits.    Because of the shorter inspiral time, retrograde orbits are also seen to plunge from orbits with larger eccentricity than prograde orbits.   In all cases, the larger the initial $p$, the longer the inspiral, and hence the smaller the final eccentricity.   The residual values of $e$ are plotted against $p_i$ for various inclinations in the left panel of Fig.\ \ref{fig:efinal}.   The right panel of Fig.\ \ref{fig:efinal} shows the number of orbits to plunge vs. the initial $p$, for $e_i = 0.999$.   The curves for $i = 135^{\rm o}$ and $180^{\rm o}$ drop precipitously to zero at values of $p_i$ that are already at the critical value for immediate plunge.

\subsection{Inclined orbits and the late-time flux of gravitational waves}
\label{sec:flux}

Here we study the effect of orbital inclination on the flux and frequency of gravitational waves near the end-point of the orbital evolution.  It is not our intent to provide accurate predictions for the flux itself -- in the highly relativistic regime near the onset of plunge, this may be pushing the  PN approximation beyond its regime of validity.  However, our PN results may give useful insights into the dependence of the gravitational waves on the inclination of the orbit.  To investigate this we estimate the flux as 
\begin{equation}
{\cal F} = \frac{2\pi}{P}  \left ( \frac{dE}{dp} \frac{dp}{d\theta}_{\rm RR} +  \frac{dE}{de} \frac{de}{d\theta}_{\rm RR} \right ) \,,
\end{equation}
where $E$ is given by Eq.\ (\ref{eq4:Enew}) multiplied by $\eta M$,  $P$ is given by (\ref{eq3:Pnew}), and ${dp}/{d\theta}_{\rm RR}$ and ${de}/{d\theta}_{\rm RR}$ are given by (\ref{eq:dpdthetaRRnew}) and (\ref{eq:dedthetaRRnew}).  Since the radiation-reaction expressions do not include $\chi^2$ terms, we set $\chi^2 = 0 $ in the expressions for $E$ and $P$.  The result is 
\begin{widetext}
\begin{eqnarray}
{\cal F} &=& {\cal C}^{-1} \frac{32\eta}{5} \frac{c^5}{G} \left (\frac{GM}{c^2 p} \right )^5 (1-e^2)^{3/2} \biggl \{ \left (1 + \frac{73}{24} e^2 + \frac{37}{96} e^4 \right ) - \frac{GM}{c^2 p} \left ( \frac{95216+306240 e^2+53242 e^4+ 715 e^6}{5376} \right )
\nonumber \\
&& \quad 
- \left (\frac{GM}{c^2 p} \right )^{3/2} \chi \cos \iota \left ( \frac{1936 + 12024 e^2+ 6582 e^4+195 e^6}{ 192} \right )
\nonumber \\
&& \quad 
+ \left (\frac{GM}{c^2 p} \right )^{2} \left ( \frac{  121274560  + 421538216 e^2 +84768510 e^4 - 1355193 e^6  - 659322 e^8}{580608} \right ) \biggr \} \,,
\end{eqnarray}
where
\begin{equation}
{\cal C} = 1 + \frac{3}{8} \frac{GM}{c^2 p} (16 - 5e^2) +  6 \left (\frac{GM}{c^2 p} \right )^{3/2} \chi \cos \iota
 - \frac{3}{128} \left (\frac{GM}{c^2 p} \right )^2 \left [ 448 - 88 e^2 + 35e^4 - 320 (1-e^2)^{3/2}  \right ] \,.
\end{equation}
\end{widetext}
We estimate the gravitational-wave frequency as twice the orbital frequency, or $\omega_{\rm GW} = 4\pi/P$.   
We then evaluate the flux and frequency at the point of plunge for various orbital evolutions.  Both quantities depend almost entirely on the value of $p$ at plunge and on the orbital inclination and $\chi$.  Because the eccentricity is generally less than $0.5$ in all cases, the dependence on $e$ is very weak, and therefore largely independent of the initial values of $p$ and $e$.  The top panel of Fig.\ \ref{fig:fluxvsinc}  shows the flux for $p_i = 100 GM/c^2$ and $e_i = 0.999$, as a function of $\chi$, for a range of inclinations, all normalized to the flux for $\iota = 0$, {\em i.e.} for equatorial prograde orbits. 
For the GW frequency, we normalize to the frequency for $\chi = 0$, so that the true GW frequency is given by 
\begin{equation}
\omega_{\rm GW} = 0.057  \,\left (\frac{10^6 M_\odot}{M} \right ) g(\chi,\iota) \,{\rm Hz}\,,
\end{equation}
where $g(\chi,\iota)$ is the function plotted in the right panel of Fig.\ \ref{fig:fluxvsinc}  .

The strong suppression of the energy flux for highly inclined and retrograde orbits is mostly due to the fact that inclined orbits plunge from greater distances than do equatorial prograde orbits.  For example, for $\chi =0.6$, the factor of 140 suppression of the flux between $0^{\rm o}$  and $180^{\rm o}$ inclination is accounted for within a factor of two by the factor of $2.3$ between the values of $p$ at plunge for the two types of orbit (see Fig.\ \ref{fig:pcapture}), raised to the fifth power.   We have not plotted the corresponding curve for  $\chi =1$;  while the PN expansion should be a reasonable approximation for the flux for retrograde orbits, with $p$ between $8$ and  $15 GM/c^2$, it is not likely to be reasonable for the prograde equatorial orbit, with $p \sim 3 GM/c^2$.      

\subsection{Evolution in terms of time} 
\label{sec:evolutiontime}

\begin{figure*}[t]
\begin{center}

\includegraphics[width=3.5in]{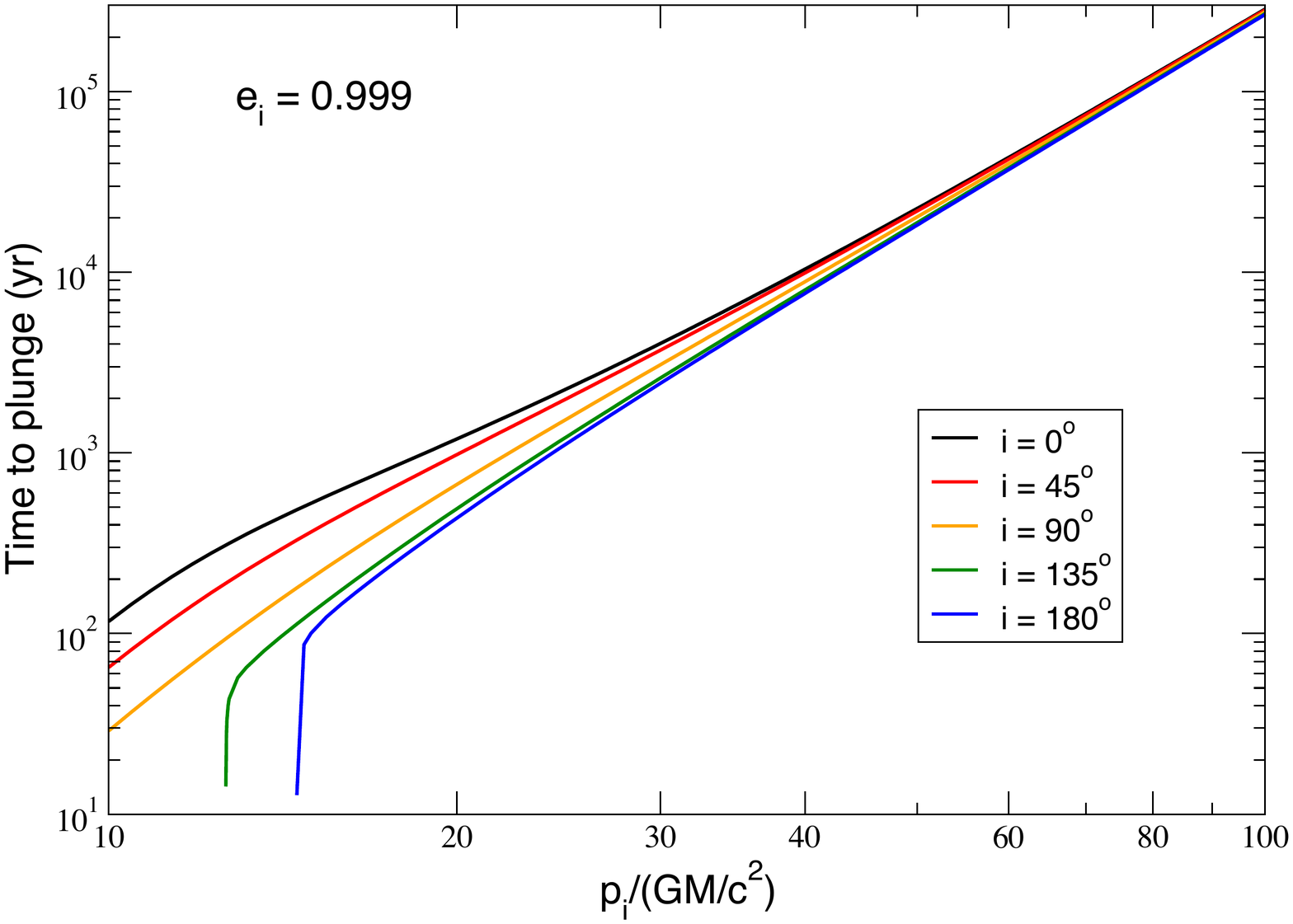}
\includegraphics[width=3.5in]{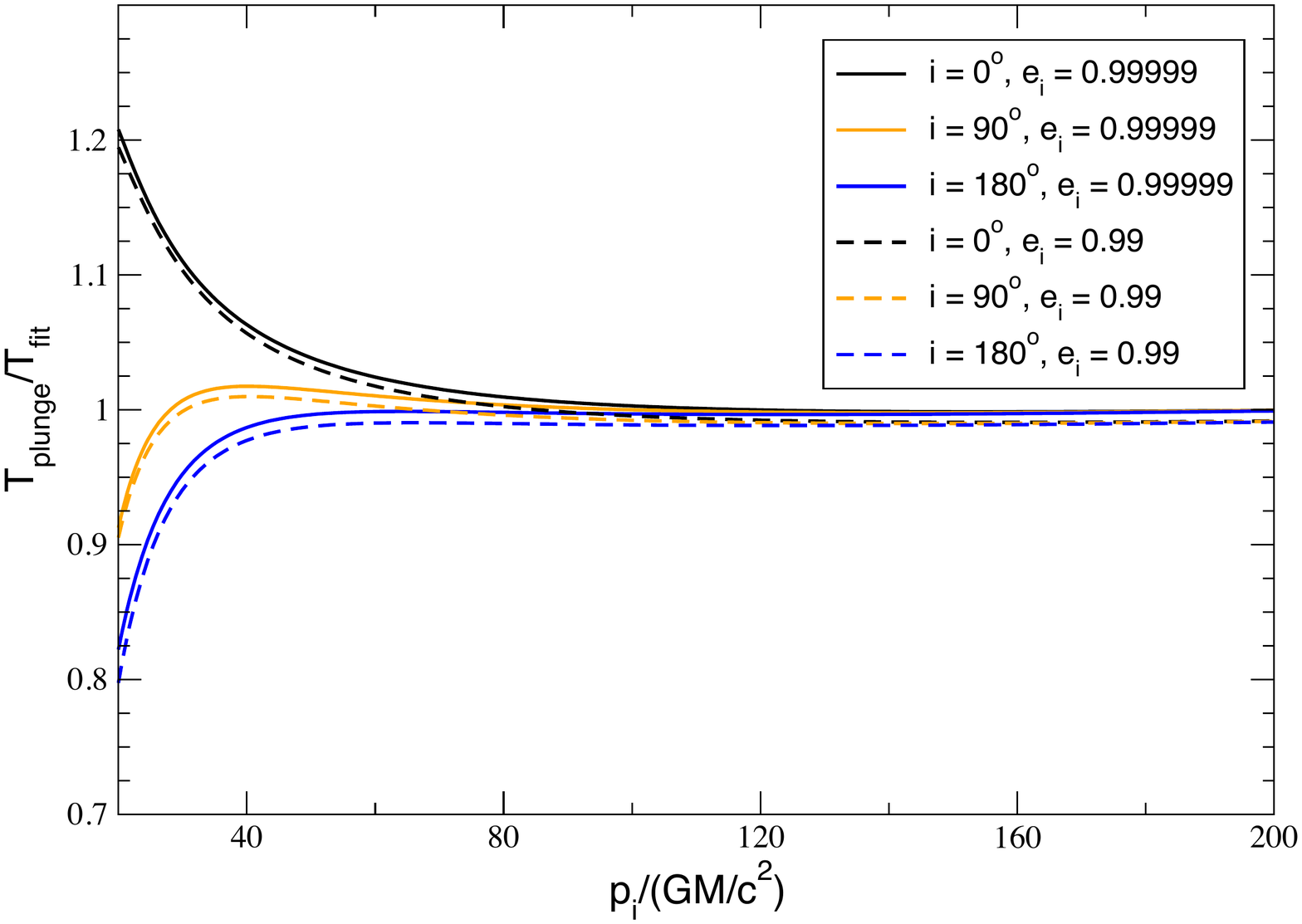}

\caption{\label{fig:timetoplunge} Left: time to plunge in years vs. initial $p$ for a $10^6 \, M_\odot$ black hole; $e_i = 0.999$ and $\chi =1$. Right: comparison between calculated plunge time and analytic fit of Eq.\ (\ref{eq:Tplungefit}); for $p$ from $20$ to $200 GM/c^2$, and $\chi =1$. }
\end{center}
\end{figure*}

We now address the evolution of our orbits with respect to time.  We assume that, for each orbit element $\tilde{X}_\alpha$, 
we can approximate
\begin{equation}
\frac{d \tilde{X}_\alpha}{dt} = \frac{2\pi}{P}  \frac{d \tilde{X}_\alpha}{d\theta}  \,.
\label{eq:dXdttimetoplunge}
\end{equation}
This is not strictly the same as having solved the original Lagrange planetary equations (\ref{eq2:Lagrange}) in terms of time directly, since time-averaged orbit elements are not the same as angle-averaged elements, beyond first order in perturbation theory.  But since the differences occur at higher PN orders, Eq.\ (\ref{eq:dXdttimetoplunge}) will serve our present purposes.   Figure \ref{fig:pevstime} shows the evolution of $p$ and $e$ vs.\ time for various initial values of $p$, for a $10^6 M_\odot$ black hole, and the left panel of Figure \ref{fig:timetoplunge} shows the time to plunge as a function of $p_i$ for various inclinations.   Because the initial periods for these highly eccentric orbits are so long for a given $p$, a relatively small number of orbits consumes a great deal of time, so that nothing dramatic happens until very late (compare Figs.\ \ref{fig:pevstime} and \ref{fig:pevsphase}).

Because of this fact, it is possible to derive a simple analytic formula that encapsulates the time to plunge to reasonable accuracy, and then to refine it empirically to obtain a formula that also accounts for the differences due to spin-orbit effects.  We first note from Eqs.\ (\ref{eq:dpdthetaRRnew}) and (\ref{eq:dedthetaRRnew}) at leading quadrupole order, and for $e \approx 1$, that 
\begin{equation}
\frac{dp}{de} \approx \frac{72}{85} \frac{p}{e} \,,
\end{equation}
so that $e \approx e_i (p/p_i)^{85/72}$, as long as $e \sim 1$.   Then, again to quadrupole order,
\begin{equation}
\frac{dp}{dt} \approx \frac{2\pi}{P} \frac{dp}{d\theta}  = - 24 \eta c \left ( \frac{GM}{c^2 p} \right )^3 (1-e^2)^{3/2} \,,
\end{equation}
where we have used the Newtonian approximation for $P$.   Substituting $x = p/p_i$ and $e = e_i x^{85/72}$,  we obtain
\begin{equation}
\frac{dx}{dt} = - 24 \eta \left ( \frac{c^3}{GM} \right ) \left ( \frac{GM}{c^2 p_i} \right )^4 \frac{(1-e_i^2 x^{85/36})^{3/2}}{x^3} \,.
\label{eq:dxdtplunge}
\end{equation}
The time to plunge is then given by inverting Eq.\ (\ref{eq:dxdtplunge}) and integrating over $x$ from $x_{\rm plunge}$ to unity.  Since $x_{\rm plunge} \ll 1$ and since the integrand varies as $\sim x^3$ for small $x$, we can safely integrate over $x$ from $0$ to $1$, to obtain
\begin{equation}
T_{\rm plunge} = \frac{1}{96 \eta} \left ( \frac{GM}{c^3} \right )\epsilon^{-4} G(e_i) \,,
\label{eq:Tplunge0}
\end{equation}
where $\epsilon = GM/c^2 p_i$, and 
\begin{eqnarray}
G(e_i) &=& 4 \int_0^1  \frac{x^3 dx}{(1-e_i^2 x^{85/36})^{3/2}} 
\nonumber \\
&=&   {_2F_1} \left ( \frac{3}{2}, \frac{144}{85} ; \frac{229}{85} ;e_i^2 \right ) 
\nonumber \\
&=&  -\frac{11328}{935} \sqrt{\pi} \frac{\Gamma (59/85)}{\Gamma(33/70)} \,{_2F_1} \left ( \frac{3}{2}, \frac{144}{85} ; \frac{3}{2}; Z \right )  
\nonumber \\
&&
+ \frac{288}{85} Z^{-1/2}\,  {_2F_1} \left ( 1, \frac{203}{170} ; \frac{1}{2}; Z \right ) \,,
\label{eq:G}
\end{eqnarray}
where $_2F_1 (a,b;c;z)$ is the hypergeometric function and $Z = 1-e_i^2$.  Thus for a chosen $\epsilon$ and $e_i$, $T_{\rm plunge}$ scales linearly with the mass of the black hole.  The final representation of $G(e_i)$ is the most useful for values of $e_i$ near unity; it can be expressed as a series expansion of the form
\begin{equation}
G(e_i) \approx  \frac{288/85}{\sqrt{1-e_i^2}} -5.925739 + 8.091903 \sqrt{1-e_i^2}  + \dots \,.
\label{eq:G2}
\end{equation}
Including only the first term in Eq.\ (\ref{eq:G2}), we obtain $T_{\rm plunge} = (3/85\eta)(GM/c^3)\epsilon^{-4} (1-e_i^2)^{-1/2}$, which is in agreement with the results of Peters \cite{1964PhRv..136.1224P} (see also Eq.\ (26) of \cite{2009MNRAS.395.2127O}).
We will adjust the coefficients slightly from the nominal values provided by Eq.\ (\ref{eq:G2}) to get a decent fit to the actual dependence of $T_{\rm plunge}$ on $e_i$.   Including PN corrections to Eq.\ (\ref{eq:Tplunge0}) particularly to account for spin-orbit effects, and adjusting the power of $\epsilon$ slightly from $4$ to $3.96$, we obtain the fit given in Eq.\ (\ref{eq:Tplungefit}).
As shown in the right panel of Fig.\ \ref{fig:timetoplunge}, this gives a surprisingly good fit to the time to plunge, to better than a percent for $p_i$ ranging from 80 to many hundred $GM/c^2$, to a few percent for $p_i$ between 40 and $80 \, GM/c^2$ and to better than 20 percent down to $p_i = 20GM/c^2$. 
For example, for $p_i = 100$ and $e_i = 0.999$, the difference in plunge time between $\iota = 0^{\rm o}$ and $\iota = 180^{\rm o}$ is about seven percent, while Eq.\ (\ref{eq:Tplungefit}) agrees with the calculated time for each inclination to $0.3$ percent.  The agreement as a function of $e_i$ for a given $p_i$ is at the level of fractions of a percent over values of $e_i$ ranging from $0.99$ to $0.99999$ that imply a range of over 4 orders of magnitude in orbital period.   We have not attempted to tweak the formula to improve the fit for smaller values of $p$, since the orbits in this regime are already sufficiently relativistic that the PN approximation is becoming less accurate anyway.

\begin{table*}[t]
\caption{\label{tab:paucarlos} Time to plunge: a comparison}
\begin{ruledtabular}
\begin{tabular}{ccccccc}
MBH mass&MBH spin&Semimajor axis&Eccentricity&Inclination&\multicolumn{2}{c}{Time to Plunge (yr)}\\
($M_\odot$)&$\chi$&$a_i$ (pc)&$1-e_i$&$\iota$ (radians)&This work&Ref \cite{2013MNRAS.429.3155A}\\
\hline

$5.0 \cdot 10^4$&$0.30$&$9.57\cdot 10^{-3}$&$10^{-6}$&$0.0$&$93$&$392$\\
$5.0 \cdot 10^4$&$0.30$&$9.57\cdot 10^{-3}$&$10^{-6}$&$0.7$&$***$&$392$\\
$5.0 \cdot 10^5$&$0.30$&$9.57\cdot 10^{-2}$&$10^{-6}$&$0.0$&$934$&$4.31 \cdot 10^4$\\
$5.0 \cdot 10^5$&$0.30$&$9.57\cdot 10^{-2}$&$10^{-6}$&$1.0$&$***$&$4.31 \cdot 10^4$\\
$1.0 \cdot 10^7$&$0.30$&$1.91$&$10^{-6}$&$0.0$&$1.87 \cdot 10^4$&$4.02 \cdot 10^6$\\
$1.0 \cdot 10^7$&$0.30$&$1.91$&$10^{-6}$&$1.0$&$***$&$3.78 \cdot 10^6$\\
$1.0 \cdot 10^6$&$0.70$&$1.91\cdot 10^{-2}$&$10^{-5}$&$0.0$&$960$&$1.35 \cdot 10^4$\\
$1.0 \cdot 10^6$&$0.70$&$1.91\cdot 10^{-2}$&$10^{-5}$&$1.0$&$644$&$1.20 \cdot 10^4$\\
$5.0 \cdot 10^6$&$0.70$&$9.57\cdot 10^{-2}$&$10^{-5}$&$0.0$&$4.80 \cdot 10^3$&$1.55 \cdot 10^5$\\
$5.0 \cdot 10^6$&$0.70$&$9.57\cdot 10^{-2}$&$10^{-5}$&$1.0$&$3.22 \cdot 10^3$&$1.36 \cdot 10^5$\\
$5.0 \cdot 10^7$&$0.70$&$9.57\cdot 10^{-1}$&$10^{-5}$&$0.0$&$4.80 \cdot 10^4$&$1.55 \cdot 10^7$\\
$5.0 \cdot 10^7$&$0.70$&$9.57\cdot 10^{-1}$&$10^{-5}$&$1.0$&$3.22 \cdot 10^4$&$1.35 \cdot 10^7$\\
$1.0 \cdot 10^6$&$0.99$&$1.91\cdot 10^{-2}$&$10^{-5}$&$0.0$&$1.66 \cdot 10^3$&$1.43 \cdot 10^4$\\
$1.0 \cdot 10^6$&$0.99$&$1.91\cdot 10^{-2}$&$10^{-5}$&$1.0$&$782$&$1.17 \cdot 10^4$\\
$1.0 \cdot 10^7$&$0.99$&$1.91\cdot 10^{-1}$&$10^{-5}$&$0.0$&$1.66 \cdot 10^4$&$1.44 \cdot 10^6$\\
$1.0 \cdot 10^7$&$0.99$&$1.91\cdot 10^{-1}$&$10^{-5}$&$1.0$&$7.82 \cdot 10^3$&$1.18 \cdot 10^6$\\
$5.0 \cdot 10^7$&$0.99$&$9.57\cdot 10^{-1}$&$10^{-5}$&$0.0$&$8.32 \cdot 10^4$&$1.65 \cdot 10^7$\\
$5.0 \cdot 10^7$&$0.99$&$9.57\cdot 10^{-1}$&$10^{-5}$&$1.0$&$3.91 \cdot 10^4$&$1.35 \cdot 10^7$\\
\end{tabular}
\end{ruledtabular}
\end{table*} 

Amaro-Seoane {\em et al.} \cite{2013MNRAS.429.3155A}  (hereafter ASF) have also endeavored to analyse the effects of black hole spin on the evolution of orbits.  They employed a hybrid approach that combined 2PN results for the conservative orbital motion, black-hole perturbation theory results for gravitational radiation, and an adiabatic assumption that corrected orbits in response to the changes in $E$, $L_e$ and $C$ \cite{2002PhRvD..66f4005G,2006PhRvD..73f4037G,2011PhRvD..84l4060S}.   In particular, ASF reported results for the time to plunge (Table 1 of \cite{2013MNRAS.429.3155A}) for a range of black hole masses and spins, for a range of initial semimajor axes and eccentricities, and for inclinations of $0^{\rm o}$ and $57.3^{\rm o}$ (one radian).   As they point out, all the orbits discussed in that table actually correspond to the single choice $p_i = 8 GM/c^2$, corresponding to the critical value for plunge for a Schwarzschild black hole, but above the critical value for prograde orbits around a Kerr black hole (see Fig.\ \ref{fig:pcapture}, where our value for the critical $p$ for Schwarzschild is closer to $ 9 GM/c^2$).   With this common choice of initial condition for $p$, then for all cases with the same values of $e_i$, $\iota_i$ and $\chi$, the time to plunge should scale with black-hole mass [see Eq.\ (\ref{eq:Tplunge0})].   Table \ref{tab:paucarlos} shows a selection of results comparing our computation of time to plunge with those from  Table 1 of ASF.   The first thing to notice is that the scaling with mass is not apparent in the results of ASF.  In addition, our times to plunge are consistently shorter than those of ASF and the dependence on the angle of inclination is stronger in our case.  The starred entries in Table \ref{tab:paucarlos} are cases where the initial orbits are already below our plunge criterion of Eq.\ (\ref{eq:condition}), and thus the plunge time is zero.   We must acknowledge that, because $p$ in our calculations of $T_{\rm plunge}$  is evolving from $9 GM/c^2$ to around $3 GM/c^2$, the PN approximation is being pushed up to or beyond its limit of validity, so we do not wish to claim too much accuracy for our values of  $T_{\rm plunge}$ in Table \ref{tab:paucarlos}.   Nevertheless, the scaling with black-hole mass and the dependence on inclination should be robust.  Reconciling the differences between these two approaches is a subject for future research.

\section{Discussion and conclusions}
\label{sec:conclusions}

We have used post-Newtonian theory to analyze the motion of stars orbiting rotating black holes in orbits that are initially very eccentric and with arbitrary inclinations relative to the black hole's equatorial plane.  We incorporated conservative terms from the test body equations of motion in the Kerr geometry, including all spin effects through 3PN order, and radiation reaction contributions to the equations of motion through 4.5PN order, including spin-orbit contributions, and carefully solved the evolution equations for the osculating orbit elements to the corresponding PN order using a two-timescale approach.  The orbits were terminated when the Carter constant associated with the orbit dropped below a critical value.  We found that retrograde orbits terminate farther from the black hole and with larger residual orbital eccentricities than do prograde orbits, and consequently that the flux of gravitational radiation from the final stage of retrograde orbits can be substantially suppressed relative to flux from the comparable prograde orbit.    We also provided a number of results in forms that are ``ready-to-use'' in numerical simulations of star clusters orbiting a spinning central black hole, include two-body equations of motion and an analytic formula for the time to plunge  that take into account the effect of the spin of the black hole.  

A key question is how well the PN approximation can be trusted in the relativistic regimes we consider here.  The inverse of the relativistic expansion parameter $GM/c^2 p$ ranges from values as large as the hundreds when the star is injected into a high-eccentricity orbit from the surrounding cluster to as small as $3$ when a prograde orbit terminates at the plunge point.  Consequently, some caution is in order in interpreting these results.   While the PN approximation has proven to be ``unreasonably effective'' in describing the dynamics of comparable-mass binary systems \cite{2011PNAS..108.5938W}, its effectiveness is not so clear in the extreme mass ratio limit, particularly in the absence of alternative approaches such as numerical relativity or self-force theory against which detailed comparisons could be made.   It would be interesting to see if various resummations of the PN sequence, along the lines of the ``effective one body'' (EOB) approach \cite{1999PhRvD..59h4006B,2000PhRvD..62f4015B} might be useful for the eccentric orbits being considered here (see {\em e.g.} \cite{2012PhRvD..86l4012B}).

\acknowledgments

This work was supported in part by the National Science Foundation,
Grant Nos.\  PHY 13--06069 \& 16--00188.   We are grateful for the hospitality of  the Institut d'Astrophysique de Paris where much of this work was carried out.     We thank Tal Alexander, Anna Heffernan, David Merritt and Bernard Whiting for useful discussions. 

\appendix

\section{Motion of a particle around a spinning black hole in harmonic coordinates}
\label{app:Kerr}

In this Appendix, we provide some of the details underlying the equations of motion used in this paper.  Parts of the following subsection are based on unpublished work by CMW and A. Le Tiec.

\subsection{Test-body equations to 3PN order}
\label{sec:kerreqns}

The conservative equations of motion used in this paper are those of a test body orbiting a Kerr black hole, to 3PN order and in harmonic coordinates.  We begin with the Kerr line element in standard Boyer-Lindquist coordinates \cite{1967JMP.....8..265B}, given by
\begin{eqnarray}
ds^2 &=& - \biggl ( 1 - \frac{2mr_{\rm bl}}{\rho^2} \biggr ) c^2 dt_{\rm bl}^2
 + \frac{\rho^2}{\Delta} dr_{\rm bl}^2 + \rho^2 d\theta_{\rm bl}^2
 \nonumber \\
 && 
  + \biggl ( r_{\rm bl}^2 + \frac{a^2}{c^2} + \frac{2mr_{\rm bl}a^2 \sin^2 \theta_{\rm bl}}{c^2 \rho^2} \biggr ) \sin^2 \theta_{\rm bl} d\phi_{\rm bl}^2 
  \nonumber \\
 && 
 - \frac{4mar_{\rm bl}}{\rho^2} \sin^2 \theta_{\rm bl} dt_{\rm bl} d\phi_{\rm bl} \,,
 \label{eq:KerrBL}
\end{eqnarray}
where $m \equiv GM/c^2$ is the geometrized mass and $a = GM\chi /c$ is the Kerr parameter, related to the spin angular momentum $S$ by 
$a \equiv S/M$, with $|\chi | \le 1$ required for the metric to describe a black hole; $\rho^2 = r_{\rm bl}^2 + (a/c)^2 \cos^2 \theta_{\rm bl}$, and 
$\Delta = r_{\rm bl}^2  - 2mr_{\rm bl}+ (a/c)^2$.   The transformation from Boyer-Lindquist coordinates to Cartesian-like harmonic coordinates satisfying $\partial_\alpha (\sqrt{-g} g^{\alpha\beta} \partial_\beta x_{\rm H}^{(\gamma)}) =0$ is given by \cite{1997PhRvD..56.4775C,2000LRR.....3....5C,2008PhRvD..77j4001H}
\begin{eqnarray}
t_{\rm H} &=& t_{\rm bl} \,,
\nonumber \\
x_{\rm H} + i y_{\rm H} &=& (r_{\rm bl} - m + ia/c) e^{i \psi} \sin \theta_{\rm bl} \,,
\nonumber \\
z_{\rm H} &=& (r_{\rm bl} - m ) \cos \theta_{\rm bl} \,,
\label{eq:harmonic}
\end{eqnarray}
where
\begin{equation}
\psi = \phi_{\rm bl} + \frac{a}{c(r_+ - r_-)} \ln \left | \frac{r_{\rm bl} - r_+}{r_{\rm bl} + r_-} \right |
\end{equation}
and $r_\pm = m \pm \sqrt{m^2 - (a/c)^2}$.   Inverting these transformations iteratively to the requisite PN order and inserting into Eq.\ (\ref{eq:KerrBL}), we obtain the metric in harmonic coordinates to 3PN order:
\begin{eqnarray}
g_{00} &=& - \left (\frac{1 - GM/c^2 r}{1 + GM/c^2 r} \right ) - \left (\frac{GM}{c^2 r} \right )^3 \chi^2 \left [ 3 ({\bm n} \cdot {\bm e} )^2 -1 \right ]  
\nonumber \\
&&
+ 2\left (\frac{GM}{c^2 r} \right )^4 \chi^2 \left [ 4 ({\bm n} \cdot {\bm e} )^2 -1 \right ] + O(\epsilon^5) \,,
\nonumber \\
g_{0j} &=& 2 \frac{GM^2}{c^4 r^2} \chi ({\bm n} \times {\bm e} )^j \left ( 1 - \frac{GM}{c^2 r} \right ) 
+ O(\epsilon^4)\,,
\nonumber \\
g_{ij} &=& \left(1 + \frac{GM}{c^2 r} \right )^2 \left (  \delta_{ij} -n_i n_j \right )
+\left (\frac{1 + GM/c^2 r}{1 - GM/c^2 r} \right ) n_i n_j 
\nonumber \\
&&
- \left (\frac{GM}{c^2 r} \right )^3 \chi^2 \left [ 3 ({\bm n} \cdot {\bm e} )^2 -1 \right ] \delta_{ij} 
\nonumber \\
&& + 2 \left( \frac{GM}{c^2 r} \right )^3 \chi ({\bm n} \times {\bm e} )^{(i} n^{j)} + O(\epsilon^4) \,,
\label{eq:KerrH}
\end{eqnarray}
where $\bm e$ is a unit vector in the direction of the black-hole spin.
The first term in $g_{00}$ is the usual Schwarzschild piece in harmonic coordinates; the second term is the contribution of the body's quadrupole moment, and the third term is a 3PN ``cross-term'' between the monopole and quadrupole terms combined with a (spin)$^2$ term.  That the terms in the time-space component $g_{0j}$ are of order $c^{-4}$ and $c^{-6}$ rather than  $c^{-3}$ and $c^{-5}$ is a reflection of the well-known fact that for black holes, effects linear in spin are effectively 1/2 PN order higher than they would be for normal slowly rotating bodies.   In $g_{ij}$ the first two terms are again the standard Schwarzschild contribution, while the third is the quadrupole term.   The final term may be dropped: it is a pure gauge term, as can be checked by making the spatial coordinate transformation
\begin{equation}
x^j = \bar{x}^j + \frac{1}{3} \left (\frac{GM}{c^2 \bar{r}} \right )^3 \chi ({\bar{\bm x}} \times {\bm e} )^{j} \,,
\end{equation}
which maintains harmonic gauge, eliminates that term, and leaves the rest of the metric unchanged, to 3PN order.

The equations of motion for a test body can then be obtained from the geodesic equation, expressed using coordinate time as the parameter,
\begin{equation}
\frac{d^2 x^i}{dt^2} + \left ( \Gamma^j_{\beta\gamma} - \Gamma^0_{\beta\gamma} \frac{v^j}{c} \right ) v^\alpha v^\beta =0\,,
\end{equation}
where $v^\alpha \equiv (c,v^j)$, and $ \Gamma^\alpha_{\beta\gamma} $ are the Christoffel symbols computed from the metric.  The result through 3PN order is Eq.\ (\ref{eq:eom2}).   Since the Kerr metric is stationary and axisymmetric, it admits the two conserved quantities of  energy (per unit mass-energy) $\tilde{E}$ and $\bm e$-component of angular momentum per unit mass, $L_e$, given by 
\begin{equation}
\tilde{E} \equiv - u_0/c \,, \qquad  L_e \equiv u_\phi \,.
\end{equation}
where $u^\alpha \equiv dx^\alpha/d\tau$ is the four-velocity of the test particle.   It is straightforward to express these to 3PN order in harmonic coordinates using Eq.\ (\ref{eq:KerrH}).  Defining $\tilde{E} \equiv 1+{E}/c^2$, we obtain
\begin{widetext}
\begin{eqnarray}
{E} &=& \frac{1}{2} v^2 - \frac{GM}{r} +\frac{1}{8c^2} \left (3 v^4 + {12} v^2 \frac{GM}{r} +4 \frac{G^2 M^2}{r^2} \right )
\nonumber \\
&& + \frac{1}{c^4} \left (\frac{5}{16} v^6 + \frac{21}{8} v^4 \frac{GM}{r} + \frac{7}{4} v^2 \frac{G^2M^2}{r^2}
+\frac{1}{2} \dot{r}^2 \frac{G^2M^2}{r^2} - \frac{1}{2} \frac{G^3 M^3}{r^3} \right )
\nonumber \\
&&
+ \frac{1}{2c^4} \left ( \frac{GM}{r} \right )^3 \chi^2 \left [3 ({\bm n} \cdot {\bm e} )^2 -1 \right ]
- \frac{2}{c^5}  \left (\frac{G M}{r} \right )^2 v^2 \chi \frac{L_z}{r} 
\nonumber \\
&&
+ \frac{1}{c^6} \left ( \frac{35}{128} v^8 + \frac{55}{16} v^6 \frac{GM}{r} +  \frac{135}{16} v^4 \frac{G^2M^2}{r^2}+\frac{3}{4} v^2 \dot{r}^2 \frac{G^2M^2}{r^2} +\frac{5}{4} v^2 \frac{G^3 M^3}{r^3}+\frac{3}{2} \dot{r}^2 \frac{G^3 M^3}{r^3} + \frac{3}{8} \frac{G^4 M^4}{r^4} \right )
\nonumber \\
&&
-  \frac{1}{4c^6}   \left (\frac{G M}{r} \right )^3 \chi^2 \left ( 3v^2  \left [3 ({\bm n} \cdot {\bm e} )^2 -1 \right ]
+ 2 \frac{GM}{r} \left [5 ({\bm n} \cdot {\bm e} )^2 -1 \right ] \right ) \,,
\label{app:E3pn} \\
L_e &=& L_z \left \{ 1 + \frac{1}{2c^2} \left (v^2 + 6 \frac{GM}{r} \right )
+ \frac{1}{8c^4} \left (3v^4 + 28 v^2 \frac{GM}{r}+ 28 \frac{G^2M^2}{r^2} \right )
- \frac{2}{c^5}  \left (\frac{G M}{r} \right )^2 \frac{L_z}{r} \chi
\right .
\nonumber \\
&& 
\left .
\quad + \frac{1}{c^6} \left (\frac{5}{16} v^6 + \frac{33}{8} v^4 \frac{GM}{r} + \frac{45}{4} v^2 \frac{G^2M^2}{r^2}
+ \frac{1}{2} \dot{r}^2 \frac{G^2M^2}{r^2} + \frac{5}{2} \frac{G^3 M^3}{r^3} \right )
\right .
\nonumber \\
&& 
\left .
\quad
- \frac{3}{2c^6} \left ( \frac{GM}{r} \right )^3 \chi^2 \left [3 ({\bm n} \cdot {\bm e} )^2 -1 \right ]
\right \}
+ \frac{2}{c^3} \frac{G^2M^2}{r} \chi \left [ ({\bm n} \cdot {\bm e} )^2 -1 \right ] \left (1 + \frac{v^2}{2c^2} \right ) \,,
\label{app:Le3pn}
\end{eqnarray}
where $L_z \equiv ( {\bm x} \times {\bm v} ) \cdot {\bm e} $.  The geometry admits an additional constant of the motion known as the Carter constant \cite{1968PhRv..174.1559C}, given by $C_0 \equiv K_{\mu\nu} u^\mu u^\nu $, where $K_{\mu\nu}$ is the Killing tensor of the metric.  However, we will find it more useful to employ a related ``Carter'' constant, given by 
\begin{eqnarray}
C &\equiv& C_0 - (a\tilde{E})^2 + 2a\tilde{E}L_e 
\nonumber \\
&=& \rho_{\rm bl}^4 \left (u^{\theta_{\rm bl}} \right )^2 + \sin^{-2} \theta_{\rm bl} L_e^2  + a^2 \cos^2 \theta_{\rm bl} (1-\tilde{E}^2) \,.
\end{eqnarray}
This definition has the advantage that, in the Schwarzschild limit ($a \to 0$), $\sqrt{C}$ is equal to the conserved total angular momentum, and in the Newtonian limit, it is equal to $L \equiv | {\bm x} \times {\bm v}|$.   Thus this version of $C$ will have a more useful post-Newtonian expansion.  Converting to harmonic coordinates using Eqs.\ (\ref{eq:harmonic}), inserting the expressions for $\tilde{E} = 1+E/c^2$ and $L_e$,  and using the fact that, in our harmonic coordinates, $(d\theta/dt)^2 = r^{-4} (L^2 - \sin^{-2} \theta L_z^2 )$, we obtain, to 3PN order,
\begin{eqnarray}
C &=& L^2 \left \{ 1 + \frac{1}{c^2} \left ( v^2 + 6\frac{GM}{r} \right )
+ \frac{1}{c^4} \left (v^4 + 10 v^2 \frac{GM}{r}+ 16 \frac{G^2M^2}{r^2} \right )
\right .
\nonumber \\
&& 
\left .
+ \frac{1}{c^4} \left (\frac{GM}{r} \right )^2 \chi^2 \left [ 2({\bm n} \cdot {\bm e} )^2 -1 \right ]
- \frac{4}{c^5}   \left (\frac{GM}{r} \right )^2 \frac{L_z}{r} \chi
\right .
\nonumber \\
&& 
\left .
+ \frac{1}{c^6} \left (v^6 + 14 v^4 \frac{GM}{r} +47 v^2 \frac{G^2M^2}{r^2}
+  \dot{r}^2 \frac{G^2M^2}{r^2} + 26\frac{G^3 M^3}{r^3} \right )
\right .
\nonumber \\
&& 
\left .
 +\frac{1}{c^6}   \left (\frac{G M}{r} \right )^2 \chi^2 \left ( v^2  \left [2 ({\bm n} \cdot {\bm e} )^2 -1 \right ]
+ 3 \frac{GM}{r} \left [ ({\bm n} \cdot {\bm e} )^2 -1 \right ] \right )
\right \}
\nonumber \\
&& - \frac{4}{c^3} \frac{G^2M^2}{r} L_z \chi \left \{ 1+ \frac{1}{c^2} \left ( v^2 + 3\frac{GM}{r} \right ) \right \}
+ \frac{1}{c^4} \left (\frac{G M}{r} \right )^2 L_z^2 \chi^2 \left \{ 1 + \frac{1}{c^2} \left ( v^2 + 6\frac{GM}{r} \right ) \right \}
\nonumber \\
&&
- \frac{G^2M^2}{c^4} \chi^2 ({\bm n} \cdot {\bm e} )^2 \left \{ v^2 - 2\frac{GM}{r} + \frac{1}{c^2} \left (v^4 +2v^2 \frac{GM}{r} + 2 \frac{G^2 M^2}{r^2} \right ) \right \}
\nonumber \\
&&
- \frac{2}{c^4} L  \dot{r} \frac{G^2 M^2}{r} \chi^2 ({\bm \lambda} \cdot {\bm e} )({\bm n} \cdot {\bm e} )
\left \{ 1 +  \frac{1}{c^2} \left ( v^2 + 6\frac{GM}{r} \right ) \right \}
- \frac{4}{c^6} \frac{G^4M^4}{r^2} \chi^2 \left [ ({\bm n} \cdot {\bm e} )^2 -1 \right ] \,,
\label{app:C3pn}
\end{eqnarray}
where $\bm \lambda \equiv {\bm \hat{h}} \times {\bm n}$,  ${\bm \hat{h}} = {\bm L}/L$ and we use the fact that $r^2 \sin \theta \, \dot{\theta} = - h  ({\bm \lambda} \cdot {\bm e} )$.  In these expressions the terms of odd orders in $1/c$ arise from spin-orbit coupling, whose effects are always shifted by half a PN order relative to monopole or quadrupole contributions. 

\subsection{Radiation reaction terms at 4PN spin-orbit and 4.5 PN order}
\label{sec:45PNterms}

Here we display the coefficients in the 4PN spin-orbit and 4.5PN contributions to radiation reaction in Eq.\ (\ref{eq:eomRR}).  The spin-orbit terms were derived from first principles in \cite{2005PhRvD..71h4027W}; the coefficients are given by
\begin{eqnarray}
A_{\rm 4SO} &=& 120 v^2 + 280 \dot{r}^2 + 453 \frac{GM}{r} \,,
\nonumber \\
B_{\rm 4SO} &=& 87 v^2 -675 \dot{r}^2 - \frac{901}{3}  \frac{GM}{r} \,,
\nonumber \\
C_{\rm 4SO} &=& 48 v^2 + 15 \dot{r}^2 + 364 \frac{GM}{r} \,,
\nonumber \\
D_{\rm 4SO} &=& 31 v^4 - 260 v^2 \dot{r}^2 + 245 \dot{r}^4 - \frac{689}{3} v^2 \frac{GM}{r}
+537 \dot{r}^2 \frac{GM}{r} + \frac{4}{3} \left ( \frac{GM}{r} \right )^2 \,.
\end{eqnarray}
Because we are dealing with a single spinning body at rest, these coefficients are independent of the ``spin-supplementary condition'' adopted.

The 4.5PN contributions were derived by Gopakumar {\em et al.} \cite{1997PhRvD..55.6030G}, extending a method developed by Iyer and Will \cite{1993PhRvL..70..113I,1995PhRvD..52.6882I}.  The idea is to write down the most general expression for 4.5PN radiation reaction terms for a binary system of point masses (in the center-of-mass frame), and to fix the coefficients in that expression by requiring that the equations yield the proper expressions for energy and angular momentum flux, accurate to 2PN order beyond the quadrupole approximation.  The procedure does not fix the coefficients uniquely, but the remaining free parameters can be shown to be related to the freedom to make coordinate or gauge changes at the relevant 4.5PN order.    At 4.5PN order, there are 12 degrees of freedom, parametrized by $\psi_1 \dots \psi_9, \, \chi_6, \, \chi_8$ and $\chi_9$, in the notation of Ref.\ \cite{1997PhRvD..55.6030G}.  The resulting functions $E_{\rm 4.5}$ and $F_{\rm 4.5}$ in Eq.\ (\ref{eq:eomRR}) are given by
\begin{eqnarray}
E_{\rm 4.5} &=& 
  \left( {\frac {779}{168}}+3\,\psi_{{2}}-3\,\chi_{{6}} \right) {v}^{6} 
    - \left( {\frac {295}{84}} + 5\,\psi_{{2}} - 5\,\chi_{{6}} - 5\,\psi_{{4}} + 5\,\chi_{{8}} \right) {v}^{4}\dot{r}^{2}
    -9\,\psi_{{7}}\dot{r}^{6}
  \nonumber \\
  && 
+ \left( {\frac {145}{6}}-7\,\psi_{{4}}+7\,\chi_{{8}}+7\,\psi_{{7}} \right) {v}^{2}\dot{r}^{4}
+ \left( {\frac {6793}{84}}-2\,\psi_{{1}}-3\,\psi_{{2}}+3\,\chi_{{6}}+3\,\psi_{{6}}-3\,\chi_{{9}} \right) {v}^{4}\frac{GM}{r}
  \nonumber \\
  &&
  - \left( {\frac {218401}{504}} + 4\psi_{{2}} + 5\,\psi_{{4}} + 6\,\psi_{{6}} - 5\,\psi_{{8}} - 2\,\chi_{{6}} - 5\,\chi_{{8}} - 6\,\chi_{{9}} \right) {v}^{2}\dot{r}^{2}\frac{GM}{r}
    \nonumber \\
  &&
  + \left( {\frac {54161}{126}}-2\,\psi_{{4}}-7\,\psi_{{7}}-8\,\psi_{{8}} \right) \dot{r}^{4}\frac{GM}{r}
  - \left( 83 + 2\,\psi_{{3}} + 3\,\psi_{{6}} - 3\,\chi_{{9}} - 3\,\psi_{{9}} \right) {v}^{2} \frac{G^2M^2}{r^2}
    \nonumber \\
  &&  
  + \left( {\frac {83407}{252}}-2\,\psi_{{6}}-5\,\psi_{{8}}-7\,\psi_{{9}} \right) \dot{r}^{2}\frac{G^2M^2}{r^2}
 + \left( {\frac {41297}{108}}-2\,\psi_{{5}}-3\,\psi_{{9}} \right) \frac{G^3M^3}{r^3}  
  \,,
  \nonumber \\
F_{\rm 4.5} &=&
 \left( {\frac {417}{28}} - \psi_{{1}} \right) {v}^{6}
 - \left( {\frac {380}{3}} - 3\,\psi_{{1}} + 3\,\chi_{{6}} \right) {v}^{4}\dot{r}^{2}
 - \left( {\frac {485}{6}} - 7\,\chi_{{8}} \right) \dot{r}^{6}
 \nonumber \\
  &&
 + \left( {\frac {34445}{168}} + 5\,\chi_{{6}} - 5\,\chi_{{8}} \right) {v}^{2}\dot{r}^{4}
+ \left( {\frac {1859}{56}}+\psi_{{1}}-\psi_{{3}} \right) {v}^{4}\frac{GM}{r}
 \nonumber \\
  &&
   - \left( {\frac {16687}{42}} - 4\,\psi_{{1}} - 4\,\psi_{{3}} - 3\,\chi_{{6}} + 3\,\chi_{{9}} \right) {v}^{2}\dot{r}^{2}\frac{GM}{r}
\nonumber \\
  &&
   + \left( {\frac {99499}{252}} + 2\,\chi_{{6}} + 5\,\chi_{{8}} + 6\,\chi_{{9}} \right) \dot{r}^{4}\frac{GM}{r}
  - \left( {\frac {2967}{28}} - \psi_{{3}} + \psi_{{5}} \right) {v}^{2}\frac{G^2M^2}{r^2}
    \nonumber \\
  && 
 \nonumber \\
  && 
  + \left( {\frac {3166}{21}}+2\,\psi_{{3}}+5\,\psi_{{5}}+3\,\chi_{{9}} \right) \dot{r}^{2}\frac{G^2M^2}{r^2}
  + \left( {\frac {395929}{2268}}+\psi_{{5}} \right) \frac{G^3M^3}{r^3} 
\,.
\end{eqnarray}
It turns out, not surprisingly, that the arbitrary parameters completely drop out of the orbit-averaged equations for the orbital elements. 
\end{widetext}

\section{Two-scale analysis of the Lagrange planetary equations}
\label{app:twoscale}

Here we provide some of the details of the two-scale analysis that we used to find the equations for the secular evolutions of the orbit elements. 
The Lagrange planetary equations for the orbit elements $X_\alpha (t)$ take the general form
\begin{equation}
\frac{d X_\alpha (t)}{dt} = \epsilon Q_\alpha (X_\beta(t), t) \,,
\label{eq2:dXdt}
\end{equation}
where $\alpha$ labels the orbit element, $\epsilon$ is a small parameter that characterizes the perturbation, and
the $Q_\alpha (X_\beta(t), t) $ are functions of the orbit elements $X_\beta$ as well as explicit functions of $t$ (in practice, an angular variable such as the phase $\phi$ or the true anomaly $f$ is often a proxy for $t$).  Those functions are assumed to be periodic with period $P$ or $2\pi$.   To zeroth order in $\epsilon$, $dX_\alpha/dt = 0$, and thus the  $X_\alpha$ are constants. 

In general there are six orbital elements, including one related to the ``time of pericenter passage'', but for many problems for which the actual time of events is not important, it is useful to eliminate time from the problem and to work in terms of an angular variable expressing the orbital phase.  The possibilities include the phase $\phi$ measured from the ascending node, the true anomaly $f$ measured from the pericenter of the orbit, or the ``eccentric anomaly" $u$, related to $f$ by the transformations
\begin{equation}
\cos u = \frac{\cos f + e}{1+e \cos f} \,, \quad \sin u = \frac{\sqrt{1-e^2} \sin f}{1+e \cos f} \,,
\end{equation}
with $r = p/(1+e \cos f) = a (1-e \cos u)$, where $a = p(1-e^2)$.  
The transformations between  these variables and time $t$  are given by  
\begin{eqnarray}
\frac{d\phi}{dt} &=& \frac{h}{r^2}  - \cos \iota \frac{d\Omega}{dt}  \,,
\nonumber \\
\frac{df}{dt} &=& \frac{h}{r^2}  - \frac{d\omega}{dt} - \cos \iota \frac{d\Omega}{dt}  \,,
\nonumber \\
\
\frac{du}{dt} &=& \frac{\sqrt{1-e^2}}{1+e \cos f} \frac{df}{dt} \,.
\end{eqnarray}
Another possibility is the ``mean anomaly'', $\ell \equiv 2\pi t/P$, where $P$ is the orbital period.
After one of these transformations, we arrive at five orbit element equations in the generic form
\begin{equation}
\frac{d X_\alpha (\phi)}{d\phi} = \epsilon Q_\alpha (X_\beta(\phi), \phi) \,,
\label{eqapp:dXdf}
\end{equation}
where $Q_\alpha (X_\beta(\phi), \phi) = (dt/d\phi) Q_\alpha (X_\beta(t), t)$, where we use $\phi$ to represent generically the phase variable $\{ \phi, \, f ,\, u,\, \ell \}$ being used.

As outlined in Sec.\ \ref{sec:twoscale}, we define the long timescale variable $\theta \equiv \epsilon \phi$,
write the derivative with respect to $\phi$ as
${d}/{d\phi} \equiv \epsilon {\partial}/{\partial \theta} + {\partial}/{\partial \phi}$
and define
\begin{equation}
X_\alpha (\theta, \phi) \equiv \tilde{X}_\alpha (\theta) + \epsilon Y_\alpha (\tilde{X}_\beta (\theta), \phi) \,.
\label{eqapp:ansatz}
\end{equation}
where $\tilde{X}_\alpha (\theta)$ is the average of $X_\alpha$ over $\phi$, and $ Y_\alpha$ is the average-free part, 
where the average and average-free parts are defined by 
\begin{equation}
\langle A \rangle \equiv \frac{1}{2\pi} \int_0^{2\pi} A(\theta,\phi) d\phi \,,  \quad  {\cal AF}(A) \equiv  A(\theta,\phi) - \langle A \rangle \,, 
\label{eqapp:averagedef}
\end{equation}
where the integrals are done holding $\theta$ fixed.  

Substituting our definition of $X_\alpha$ into  Eq.\ (\ref{eqapp:dXdf}), and taking the average and average-free parts, we obtain the two main equations of the procedure
\begin{subequations}
\begin{align}
\frac{d\tilde{X}_\alpha}{d\theta} &= \langle Q_\alpha (\tilde{X}_\beta + \epsilon Y_\beta, \phi) \rangle \,,
\label{eqapp:aveq}\\
\frac{\partial Y_\alpha}{\partial \phi} &= {\cal AF} \left (Q_\alpha (\tilde{X}_\beta + \epsilon Y_\beta, \phi) \right )  - \epsilon \frac{\partial Y_\alpha}{\partial \tilde{X}_\gamma} \frac{d\tilde{X}_\gamma}{d\theta} \,.
\label{eqapp:avfreeeq}
\end{align}
\label{eqapp:maineq}
\end{subequations}
Note that, by virtue of our assumption that $\theta$ and $\phi$ are independent, $\partial Y_\alpha/\partial \tilde{X}_\gamma$ is automatically average-free.

In integrating Eq.\ (\ref{eqapp:avfreeeq}), it is useful to consider the general equation
\begin{equation}
\frac{\partial A}{\partial \phi} = {\cal AF} (B) \,,
\end{equation}
for some functions $A$ and $B$, where $B$ is periodic with period $2\pi$.   The solution is 
\begin{equation}
A = \int_0^\phi {\cal AF}(B) d\phi' + C \,,
\label{eq2:Ageneral}
\end{equation}
where $C$ is a constant of integration, fixed by the condition that $\langle A \rangle = 0$.  Note that for any function $D$,
\begin{align}
\left \langle \int_0^\phi D d\phi' \right \rangle &= \frac{1}{2\pi} \int_0^{2\pi} d\phi \int_0^\phi D \, d\phi'
\nonumber \\
&= 2\pi \langle D \rangle - \langle \phi D \rangle \,,
\label{eq2:averageintegral}
\end{align} 
after switching the order of integration.   It is then simple to show that $C = \langle \phi B \rangle - \pi \langle B \rangle$, so that the general, average-free solution of Eq.\ 
(\ref{eq2:Ageneral}) is
\begin{equation}
A = \int_0^\phi B \,d\phi' - (\phi+ \pi) \langle B \rangle + \langle \phi\,B \rangle \,.
\label{eq2:Asolution}
\end{equation}
Note that $\langle \partial A /\partial \phi \rangle = 0$, and that $A(0) = A(2n\pi) = \langle (\phi-\pi) B \rangle$, for any integer $n$, thus $A$ is also periodic.   Averages of expressions involving integrals satisfy the useful property
\begin{equation}
\langle A \int_0^\phi B d\phi' \rangle = - \langle B \int_0^\phi A d\phi' \rangle + 2\pi \langle A \rangle\langle B \rangle \,.
\end{equation}

We now iterate Eqs.\ (\ref{eqapp:maineq}) in powers of $\epsilon$.  We first expand
\begin{equation}
Q_\alpha (\tilde{X}_\beta + \epsilon Y_\beta, \phi) \equiv \sum_{m=0}^\infty \frac{\epsilon^m}{m!} Q^{(0)}_{\alpha,\beta \dots \gamma} Y_\beta \dots Y_\gamma \,,
\end{equation} 
where we sum over repeated indices $\beta$, $\gamma$, etc, and where
\begin{align}
Q_\alpha^{(0)} &\equiv Q_\alpha (\tilde{X}_\beta, \phi) \,,
\\
Q_{\alpha,\beta \dots \gamma}^{(0)} & \equiv \frac{\partial^m Q_\alpha^{(0)} }{\partial \tilde{X}_\beta \dots \partial \tilde{X}_\gamma } \,.
\end{align}
We also expand 
\begin{equation}
Y_\alpha  \equiv Y^{(0)}_\alpha + \epsilon Y^{(1)}_\alpha + \epsilon^2 Y^{(2)}_\alpha + \dots \,.
\end{equation}
Then, to order $\epsilon^0$, we have
\begin{equation}
\frac{d\tilde{X}_\alpha}{d\theta} = \langle Q_\alpha^{(0)} \rangle \,,
\label{dXdtheta0}
\end{equation}
and
\begin{equation}
\frac{dY_\alpha^{(0)}}{d\phi} = {\cal AF} \left (Q_\alpha^{(0)} \right ) \,,
\end{equation}
with the solution
\begin{equation}
Y_\alpha^{(0)} = \int_0^\phi Q_\alpha^{(0)} d\phi' - (\phi+\pi) \langle Q_\alpha^{(0)} \rangle + \langle \phi \,Q_\alpha^{(0)} \rangle \,.
\label{eq2:Y0sol}
\end{equation}
At first order in $\epsilon$, we find, from Eq.\ (\ref{eqapp:avfreeeq}) 
\begin{equation}
\frac{dY_\alpha^{(1)}}{d\phi} = {\cal AF} \left ( Q_{\alpha,\beta}^{(0)}  Y_{\beta}^{(0)} \right ) - Y_{\alpha,\beta}^{(0)} \frac{d\tilde{X_\beta}}{d\theta} \,.
\end{equation}
Substituting for $Y_{\alpha}^{(0)}$ from Eq.\ (\ref{eq2:Y0sol}) and for $d\tilde{X}_\beta/d\theta$ from Eq.\ (\ref{dXdtheta0}), and making use of (\ref{eq2:Asolution}), we obtain
\begin{align}
Y_{\alpha}^{(1)} &= \int_0^\phi Q_{\alpha,\beta}^{(0)} d\phi' \int_0^{\phi'} Q_\beta^{(0)} d\phi'' 
\nonumber \\
& \quad + \left [ \langle \phi\,Q_\beta^{(0)} \rangle - \left (\phi+\pi \right ) \langle Q_\beta^{(0)} \rangle \right ] \int_0^\phi Q_{\alpha,\beta}^{(0)} d\phi' 
\nonumber \\
& \quad 
+ \left ( \phi+\pi \right )  \langle Q_\beta^{(0)} \int_0^\phi Q_{\alpha,\beta}^{(0)} d\phi' \rangle
\nonumber \\
& \quad 
+ \langle \phi\,Q_{\alpha,\beta}^{(0)} \int_0^\phi Q_\beta^{(0)} d\phi' \rangle
\nonumber \\
& \quad 
+ \frac{1}{2} \left ( \phi^2 - 4\pi^2/3 \right ) \langle Q_\beta^{(0)} \rangle \langle Q_{\alpha,\beta}^{(0)} \rangle
\nonumber \\
& \quad
-  \left (\phi+\pi \right )  \langle \phi \,Q_\beta^{(0)} \rangle \langle Q_{\alpha,\beta}^{(0)} \rangle
\nonumber \\
& \quad
- \frac{1}{2}  \langle Q_\beta^{(0)} \rangle \langle \phi (\phi+2\pi)Q_{\alpha,\beta}^{(0)} \rangle 
+ \langle \phi\,Q_\beta^{(0)} \rangle \langle \phi\,Q_{\alpha,\beta}^{(0)} \rangle \,.
\label{eq2:Y1sol}
\end{align}
It is straightforward to show explicitly that $\langle Y_{\alpha}^{(1)} \rangle = 0$.

To obtain $d\tilde{X}_\alpha/d\theta$ to order $\epsilon^2$, we expand Eq.\ (\ref{eqapp:aveq}) to second order in $\epsilon$ and substitute Eqs.\ (\ref{eq2:Y0sol}) and (\ref{eq2:Y1sol}) for $Y_\alpha^{(0)}$  and $Y_\alpha^{(1)}$, to obtain an expression purely in terms of derivatives and averages of $Q_\alpha^{(0)}$.  Converting back to the unscaled $\phi = \theta/\epsilon$, we obtain
\begin{align}
\frac{d\tilde{X}_\alpha}{d\phi} &= \epsilon \langle Q_\alpha^{(0)} \rangle + \epsilon^2 \biggl [ \langle Q_{\alpha,\beta}^{(0)} \int_0^\phi Q_{\beta}^{(0)} d\phi' \rangle
+  \langle Q_{\alpha,\beta}^{(0)} \rangle \langle \phi\,Q_\beta^{(0)} \rangle 
\nonumber \\
& \quad
-  \langle \phi\,Q_{\alpha,\beta}^{(0)} \rangle \langle Q_\beta^{(0)} \rangle 
- \pi \langle Q_{\alpha,\beta}^{(0)}  \rangle \langle Q_\beta^{(0)} \rangle  \biggr ] 
\nonumber \\
& \quad
+ \epsilon^3 \biggl [ \langle Q^{(0)}_{\alpha,\beta} Y^{(1)}_\beta \rangle + \frac{1}{2}  \langle Q^{(0)}_{\alpha,\beta\gamma} Y^{(0)}_\beta Y^{(0)}_\gamma \rangle \biggr ]
+ O(\epsilon^4)\,,
\label{eqapp:dXdtfinal}
\end{align}
where $Y^{(0)}_\beta$ and $Y^{(1)}_\beta$ are given by Eqs.\ (\ref {eq2:Y0sol}) and (\ref {eq2:Y1sol}).
The first term in Eq.\ (\ref{eqapp:dXdtfinal}) is the standard lowest-order result in which ``constant'' values of the orbit elements are inserted into $Q_\alpha$ and the result averaged over one period.  The higher-order terms result from the effect of periodic terms in the orbit elements on the behavior of the $Q_\alpha$.   

For the conservative parts of  the equations of motion used in this paper, the first term is of 1PN order and higher, the second is of 2PN order and higher and the third is of 3PN order and higher; thus the terms shown are sufficient to calculate the conservative evolution of the elements through 3PN order.  For the radiation reaction evolution, the first term is of 2.5PN order and higher.  Since we are only concerned with effects linear in the mass ratio $\eta$, the higher order terms will contribute only cross terms between radiation reaction and conservative effects.  Thus the second term will be of 3.5PN order and higher, while the final term will be of 4.5PN order and higher.  Thus the terms shown will be sufficient to determine the radiation-reaction evolution through 4.5PN order.

\bibliography{refs}

\end{document}